\documentclass[preprint,nofootinbib,aps,superscriptaddress,eqsecnum]{revtex4-1} 
 \pdfoutput=1
\usepackage{epsfig} 
\usepackage{amsmath} 
\usepackage{physics}
\usepackage{amssymb}
\usepackage{graphicx}
\usepackage{color}
\usepackage{float}
\usepackage{soul,xcolor}
\usepackage[font=small]{caption}
\usepackage{braket}
\usepackage{bbold}
\usepackage{subfigure}
\usepackage{braket}
\usepackage{titlesec}
\usepackage{placeins}
\usepackage{hyperref}
\usepackage{caption}
\usepackage{array}
\usepackage{booktabs}
\usepackage{subcaption}
\usepackage{changepage}
\usepackage{gensymb}
\usepackage{trimclip}
\usepackage{accents}
\usepackage{booktabs}
\usepackage{multirow}
\usepackage{tabularx}
\usepackage{adjustbox}
\usepackage{lipsum}
\usepackage{placeins}
\usepackage{comment}
\PassOptionsToPackage{unicode}{hyperref}
\PassOptionsToPackage{naturalnames}{hyperref}

\usepackage[justification=centering]{caption}
\newcolumntype{P}[1]{>{\centering\arraybackslash}p{#1}}
\hypersetup{colorlinks,linkcolor={blue},citecolor={blue},urlcolor={blue}}

\def\bea{\begin{eqnarray}}
\def\eea{\end{eqnarray}}
\def\be{\begin{equation}}
\def\ee{\end{equation}}

\begin{document}

\title{Probing new physics scenarios using high energy events at NOvA far detector}

\author{Chinmay Bera}
\email[Email Address: ]{chinmay20pphy014@mahindrauniversity.edu.in}
\affiliation{Department of Physics, \'Ecole Centrale School of Engineering - Mahindra University, Hyderabad, Telangana, 500043, India}

\author{K. N. Deepthi}
\email[Email Address: ]{nagadeepthi.kuchibhatla@mahindrauniversity.edu.in}
\affiliation{Department of Physics, \'Ecole Centrale School of Engineering - Mahindra University, Hyderabad, Telangana, 500043, India}

\begin{abstract}
NuMI Off-axis $\nu_e$ Appearance (NOvA) experiment is an ongoing long baseline neutrino oscillation experiment. The primary channels of interest are the $\nu_e$, $\bar{\nu}_e$ appearance, $\nu_\mu$, $\bar{\nu}_\mu$ disappearance 
channels analyzed in the energy window $1< E_\nu < 4$ GeV. However, NOvA far detector sees non-trivial high energy $\nu_e$, $\bar{\nu}_e$ events in the energy range $4 < E_\nu < 20$ GeV. These high energy events provide us with an opportunity to investigate the subleading new physics scenarios. In this context, we study the sensitivity of the NOvA experiment to constrain the non-standard interaction (NSI) parameters and environmental decoherence. We observe that by including high energy events (signal + background) the degeneracy around $\epsilon_{e\tau} \sim 1.6$ can be removed throughout the $\delta_{CP}$ and $\delta_{e\tau}$ range. Further, we examine the role of signal versus beam background events in removing this degeneracy. In addition, we constrain the decoherence parameter $\Gamma$ considering events from $1<E_\nu<20$ GeV. Later, assuming the presence of decoherence in nature we obtain the allowed regions in $\theta_{23}$ and $\delta_{CP}$ plane.

\end{abstract}


\maketitle

\section{Introduction}
The discovery of the neutrino oscillation phenomenon has established that neutrinos do have a (tiny) mass. This overwhelming evidence challenges the boundaries of the standard model (SM) of particle physics and motivates us to search for a more consistent theory. In this context, the study of new physics scenarios (contributions from sub-leading effects) are crucial as they provide the explanation to beyond the standard model (BSM) physics phenomenon. The ongoing and upcoming high precision neutrino oscillation experiments play pivotal role in the quest for new physics phenomenon. In this work, we analyze the far detector (FD) simulated events at NOvA experiment in the presence of two new physics (NP) paradigms, non-standard interactions (NSIs)~\cite{Wolfenstein:1977ue,Roulet:1991sm, Hattori:2002uw} and environmental decoherence~\cite{Chang:1998ea,Benatti:2000ph,Farzan:2008zv}. Here we consider individual effect of NSI and environmental decoherence on the neutrino propagation.

Environmental decoherence is the phenomenon that arises due to the interaction of neutrinos with the stochastic environment during propagation. This interaction leads to the loss of coherence among the neutrino mass states~\footnote{This phenomenon is different from the neutrino wave-packet decoherence.}. Additionally, neutrinos could interact with earth matter via non-standard interactions (NSI) (cannot be explained in the SM scenarios). Both the phenomenon, NSI and environmental decoherence, play a sub-leading (but significant) role in the standard neutrino oscillation framework.

Several studies have discussed neutrino oscillations in the framework of NSI and decoherence. Here, we provide some references relevant to the NOvA experiment. In ref.~\cite{Denton:2020uda} authors have pointed out the tension in latest NOvA and T2K results~\cite{NOvA:2021nfi,T2K:2021xwb} could be improved considering NSI in the analysis~\footnote{Considering scalar NSI authors have updated the analysis in the appendix of ref.~\cite{Denton:2022pxt}.}. Recent analysis of NOvA data~\cite{NOvA:2021nfi} ($1 < E_\nu < 5$ GeV) in the presence of NSI have reported NSI parameters $\epsilon_{e\mu} \leq 0.3$, $\epsilon_{e\tau} \leq 0.4$ and revealed a degeneracy of coupling parameter $\epsilon_{e\tau} > 1$~\cite{NOvA:2024lti}. In ref.~\cite{Kleykamp:2022dli} authors have extended the latest NOvA standard oscillation result, by adding NSI phenomenon and shown that $\epsilon_{e\tau}$ encounters degeneracy for full range of standard CP phase $\delta_{CP} \in [0:360\degree]$ and non-standard CP phase $\delta_{e\tau} \in [0:360\degree]$. On the other hand, authors in ref.~\cite{Coelho:2017zes} have proposed environmental decoherence to explain the shift in $\theta_{23}$ from its maximal mixing in the NOvA data. In a recent analysis~\cite{DeRomeri:2023dht} authors have constrained decoherence parameters considering $1<E_\nu<5$ GeV NOvA FD events. Analyzing high energy appearance events ($1<E_\nu<12$ GeV) at NOvA FD, authors in ref.~\cite{2735647} have constrained the standard oscillation parameters and concluded the sensitivity remains almost unaffected.
In this paper we analyze events from $1<E_\nu<20$ GeV at NOvA FD to study the sensitivity to new physics scenarios namely NSI and environmental decoherence parameters.

In order to analyze the effects of NSI and decoherence at high energy we include the events from high energy tail ($4 < E_\nu < 20$ GeV) along with the events from standard analysis window of NOvA i.e., $1 < E_\nu < 4$ GeV.
We employ these events in the determination of NSI parameters $\epsilon_{e\tau}$ and $\epsilon_{e\mu}$ one at a time. Including high energy events we show that the $\epsilon_{e\tau}$ degeneracy observed in ref.~\cite{Kleykamp:2022dli} can be removed throughout the $\delta_{CP}$ and $\delta_{e\tau}$ range. Further, we examine the role of signal versus beam background events in removing this degeneracy. Furthermore, we display the sensitivity to the decoherence parameter ($\Gamma$) considering energy power-law dependency ($\Gamma \propto E^n$, where $n$ is the power-law index) and obtain the upper bounds on $\Gamma$. We compare the upper bounds obtained from $1<E_\nu<4$ GeV with bounds from high energy events ($1<E_\nu<20$ GeV). In addition, assuming decoherence in nature we illustrate the uncertainty in the determination of $\theta_{23}$ considering events from $1<E_\nu<4$ GeV and $1<E_\nu<20$ GeV.

This paper is organized as follows: in section~\ref{sec:simulation-details} we provide simulation details and analysis method. In section~\ref{sec:effect-nsi-he} we show results for effects of NSI at high energy and, in section~\ref{sec:env-deco} effects of environmental decoherence at high energy. We summarize our results in section~\ref{sec:conclusion}.

\section{Simulation details}\label{sec:simulation-details}
NOvA is an ongoing long baseline neutrino oscillation experiment that uses NuMI beam facility at Fermilab. The experiment consists of two detectors, one near detector (ND) at the Fermilab site and a far detector (FD) located at Ash River, Minnesota.  A NuMI beam (ref.~\cite{Shanahan:2021jlp}) is delivered with an off-axis angle 14.6 mrad to the FD. The FD is $\sim 810$ km away from the source and receives a narrow beam with the flux peak $\sim 1.8$ GeV. This corresponds to first oscillation peak in the atmospheric sector. During propagation from source to detector, neutrinos experience the Earth's crust density $\sim 2.84~gm/cm^3$.

We simulate the NOvA experiment using publicly available GLoBES software packages~\cite{Huber:2004ka, Huber:2007ji}. In the experimental definition file we adopt fiducial target mass 14 kt. The experiment utilizes total exposure of $13.6 \times 10^{20}$ POT for $\nu$ and $12.5 \times 10^{20}$ POT for $\bar{\nu}$ which correspond to 6 years of running in $\nu$ mode and 3 years of running in $\bar{\nu}$ mode with an hourly-averaged beam power 742 kW. We adopt energy resolution function $\sigma_E = \alpha E + \beta \sqrt{E} + \gamma$~, where $\alpha = 0.11 (0.09),~\beta = 0,~\gamma = 0$ for $e$-like ($\mu$ like) events~\cite{NOvA:2018gge,NOvA:2019cyt}. We implement these information and reproduce the standard event rates as provided in recent NOvA results~\cite{NOvA:2021nfi}. To obtain the event spectrum we use the standard parameters listed in table~\ref{table:1}.

Additionally, to calculate the numerical oscillation probability in the presence of NSI, we deploy $snu.c$~\cite{Kopp:2006wp,Kopp:2007ne} extension with GLoBES. On the other hand, we incorporate a $new~probability~engine$ (based on open quantum system) in GLoBES to implement the effect of environmental decoherence.

\subsection{Analysis}\label{sec:analysis}
For our analysis purpose we include the events from $\nu_\mu$ ($\bar{\nu}_\mu$) $\rightarrow$ $\nu_e$ ($\bar{\nu}_e$) and $\nu_\mu$ ($\bar{\nu}_\mu$) $\rightarrow$ $\nu_\mu$ ($\bar{\nu}_\mu$) channels as these are the relevant channels to achieve scientific goals at NOvA experiment. We generate true simulated data and test events assuming different hypotheses (in details while present the results); then, we compare them statistically. We mention the details of true and theoretical hypotheses while discuss the respective results.
To perform statistical analysis we incorporate Poisson $\chi^2$ function which is defined as
\begin{equation}
    \chi^2=\min _{\alpha_s,\alpha_b} \sum_{\text {channels }} 2 \sum_i\left[N_i^{\mathrm{test}} - N_{i}^{\mathrm{true}}+N_{i}^{\mathrm{true}} \log \left(\frac{N_{i}^{\mathrm{true}}}{N_{i}^{\mathrm{test}}}\right)\right] + \alpha_s^2 + \alpha_b^2~,
    \label{eq:chi-sq}
\end{equation} 
where, $N_i^{\mathrm{test}}$ represents the number of test events and 
and $N_i^{\mathrm{true}}$ the true events (signal \& background) in $i$-th bin. $\alpha_s$ and $\alpha_b$ are refer to the signal and background normalization errors. We incorporate systematic uncertainties using $\emph{pull method}$~\cite{Fogli:2002pt, Huber:2002mx}.

\begin{table}[!htbp]\centering
\begin{tabular}{ |c|c|c| } 
 \hline\hline
 Parameters & True values & Test ranges \\[0.5ex]
 \hline
 $\sin^2{\theta_{12}}$ & $0.307$ & Fixed \\ 
 $\sin^2{\theta_{13}}$ & $0.021$ & $[0.02 : 0.02405]$ \\ 
 $\sin^2{\theta_{23}}$ NH (IH) & $0.57$ ($0.56$) & $[0.38 : 0.64]$ \\
 $\delta_{CP}$ NH (IH) & $0.82\pi$ ($1.52\pi$) & $[0 : 2\pi]$ \\
 $\frac{\Delta m^2_{21}}{10^{-5}~eV^2}$ & $7.53$ & Fixed \\ 
 $\frac{\Delta m^2_{32}}{10^{-3}~eV^2}$ NH (IH) & $2.41$ ($-2.45$) & $[\pm 2.29 : \pm 2.54]$ \\
 \hline\hline
\end{tabular}
 \caption{True oscillation parameters and $3\sigma$ ranges have been considered in our analysis are taken from refs.~\cite{NOvA:2021nfi,ParticleDataGroup:2018ovx}.}
 \label{table:1}
\end{table}

\section{Constraining NSI parameters with $\nu_e / \overline{\nu}_e$ high energy events}\label{sec:effect-nsi-he}

Neutrinos interact through weak force as described by the standard model. However, the interactions of neutrinos beyond the SM at a sub-leading level are not completely ruled out. The effect of these non-standard interactions (NSIs) on neutrino oscillation was first introduced by Wolfenstein in ref.~\cite{Wolfenstein:1977ue}. Later on, NSIs have widely been studied in the facet of neutrino physics (see reviews~\cite{Farzan:2017xzy,Biggio:2009nt,Ohlsson:2012kf,Miranda:2015dra,Proceedings:2019qno} and references therein).
NSI can modify neutrino production, propagation, and detection, leading to deviations from standard oscillation phenomena.
In particular, neutrino propagation is affected by neutral current (NC) NSI~\footnote{NSI due to charge current (CC) interaction of neutrinos with matter affects the production and detection of neutrinos. In this study, we exclude such impacts, because model independent bounds on the production and detection NSI are typically stronger than the propagation NSI by $\mathcal{O}(1)$.}. The new interactions are parameterized in terms of diagonal propagation NSI parameters $\epsilon_{\alpha\alpha}$ and off-diagonal parameters $\epsilon_{\alpha\beta}$ ($= |\epsilon_{\alpha\beta}|e^{i\delta_{\alpha\beta}}$, where $\delta_{\alpha\beta}$ is the non-standard CP phase.).

In this section, considering the impact of NSIs on neutrino propagation, we give an overview of the mathematical formulation of oscillation probability. Taking into account one non-zero off-diagonal NSI parameter among $\epsilon_{e\tau}$ and $\epsilon_{e\mu}$ at a time (neglecting others~\footnote{Appearance channels are significantly affected by the $\epsilon_{e\tau}$ and $\epsilon_{e\mu}$.}), we present modified oscillation probabilities and event spectra at the NOvA FD. Further, we analyze the two-dimensional sensitivity by projecting $\chi^2$ in $|\epsilon_{e\tau}|$ (or $|\epsilon_{e\mu}|$) versus standard CP-phase $\delta_{CP}$ plane and $|\epsilon_{e\tau}|$ (or $|\epsilon_{e\mu}|$) versus non-standard phase $\delta_{e\tau}$ (or $\delta_{e\mu}$) plane in both the energy window $1<E_\nu<4$ GeV and $1<E_\nu<20$ GeV.

\subsection{Formulation of oscillation probabilities in the presence of NSI}
The effective Lagrangian in the case of NC-NSI described by the dimension-six four-fermion operators is in the following form
\begin{equation}
    \mathcal{L}_{NC-NSI} = -2\sqrt{2}G_F\epsilon_{\alpha\beta}^{fC}\left(\bar{\nu}_\alpha \gamma^\mu P_L \nu_{\beta}\right) \left(\bar{f}\gamma_\mu P_C f\right).
\end{equation}
Here, $G_F$ is the Fermi constant and $\epsilon_{\alpha\beta}^{fC}$ are the NSI parameters, with $\alpha, \beta = e, \mu, \tau$; $f = e, u, d$ and $C = L, R$.
The effective Hamiltonian
\begin{equation}
    H = \frac{1}{2E}\left[U diag(0,\Delta m_{21}^2,\Delta m_{31}^2)U^\dagger + diag(A,0,0) + A\epsilon^m\right],
    \label{eq:eff-hamiltonian}
\end{equation}
where, $\Delta m^2_{jk}$ ($j,k = 1,2,3$) are the mass-squared differences, $U$ is the standard PMNS matrix, $A$ is the constant matter potential. $\epsilon^m$ represents NSI matrix ($3\times 3$) including non-standard matter interaction with elements $\epsilon_{\alpha\beta} \equiv \sum_{f,C} \epsilon_{\alpha\beta}^{fC} \frac{N_f}{N_e}$, where $N_f$ and $N_e$ denote the number density of $f(= e,u,d)$ fermion and electron in the earth matter through which neutrinos propagate. The diagonal elements of $\epsilon^m$ are real and off-diagonal elements carry a phase term $\delta_{\alpha\beta}$ in the following form $\epsilon_{\alpha\beta} = \left|\epsilon_{\alpha\beta}\right|e^{i\delta_{\alpha\beta}}$.

After diagonalizing, the effective Hamiltonian in eq.~(\ref{eq:eff-hamiltonian}) takes the form
\begin{equation}
    \tilde{H} = \frac{1}{2E}\tilde{U} diag \left(0,\Delta\tilde{m}_{21}^2,\Delta\tilde{m}_{31}^2\right) \tilde{U}^\dagger
    \label{eq:diag-hamiltonian},
\end{equation}
with $\Delta\tilde{m}_{jk}^2$ and $\tilde{U}$ are the effective mass-squared differences and modified PMNS matrix respectively.
Considering these effects the modified neutrino oscillation probability can be written as
\begin{equation}
\begin{aligned}
        P_{\alpha \beta}(L) &= \delta_{\alpha \beta} - 2\sum_{j > k} Re \left( \tilde{U}_{\beta j} \tilde{U}_{\alpha j}^* \tilde{U}_{\alpha k} \tilde{U}_{\beta k}^* \right) + 2\sum_{j > k} Re \left( \tilde{U}_{\beta j} \tilde{U}_{\alpha j}^* \tilde{U}_{\alpha k} \tilde{U}_{\beta k}^* \right)  \cos(\frac{\tilde{\Delta}m_{jk}^2}{2E}L) \\& + 2\sum_{j > k} Im \left( \tilde{U}_{\beta j} \tilde{U}_{\alpha j}^* \tilde{U}_{\alpha k} \tilde{U}_{\beta k}^* \right)  \sin(\frac{\tilde{\Delta}m_{jk}^2}{2E}L).
        \end{aligned}
        \label{eq:Prob}
\end{equation}
The expression for $\tilde{U}$ and $\Delta\tilde{m}_{jk}^2$ have been obtained considering non-zero $\epsilon_{e\tau}$ and $\epsilon_{e\mu}$ in the appendix of ref.~\cite{Bera:2025ayt}. 


\subsection{Oscillation probabilities and event rates in the presence of \texorpdfstring{$\epsilon_{e\tau}$}~}

In this subsection, we extensively discuss the effect of non-zero $\epsilon_{e\tau}$ (assuming all other NSI parameters to be zero) on the $\nu_e~(\bar{\nu}_e)$-appearance channel and event rates.
The approximate $\nu_\mu$ appearance probability considering series expansion in small parameters $\eta \equiv \frac{\Delta m^2_{21}}{\Delta m^2_{31}}, s_{13}, \epsilon_{e\tau}$ up to the first order as shown below
\begin{equation}\label{eq:app-prob-approx-eps_etau}
    \begin{aligned}
        P_{\mu e} =& \frac{2}{\hat{A} - 1}\left[\frac{\eta s_{12} c_{12} s_{13} s_{23} c_{23}}{\hat{A}}\left\{1 - \cos{(\frac{\Delta m^2_{21}L}{2E})} - \cos{(\frac{\Delta m^2_{31}L}{2E})}
         + \cos{(\frac{\Delta m^2_{32}L}{2E})}  \right\} + \right. \\
         &\left. \eta s_{12} c_{12} s_{23} c_{23}^2 \epsilon_{e\tau} \left\{-\frac{2-\hat{A}}{\hat{A}} \cos{(\frac{\Delta m^2_{21}L}{2E})}  
         - \cos{(\frac{\Delta m^2_{31}L}{2E})} + \cos{(\frac{\Delta m^2_{32}L}{2E})} + \frac{2-\hat{A}}{\hat{A}} \right\}  \right.\\ 
         &\left. - s_{13} s_{23}^2 c_{23} \epsilon_{e\tau} 
          \left\{ - \cos{(\frac{\Delta m^2_{21}L}{2E})} + \frac{\hat{A} + 1}{\hat{A} - 1} \cos{(\frac{\Delta m^2_{31}L}{2E})} + \cos{(\frac{\Delta m^2_{32}L}{2E})} - \frac{\hat{A} + 1}{\hat{A} - 1} \right\} \right.\\ 
        &\left. - s_{23}^2 c_{23}^2 \epsilon_{e\tau}^2 \left\{ - \cos{(\frac{\Delta m^2_{21}L}{2E})} + \frac{\hat{A}}{\hat{A} - 1}\cos{(\frac{\Delta m^2_{31}L}{2E})}  
         + \hat{A}\cos{(\frac{\Delta m^2_{32}L}{2E})} - \frac{1}{\hat{A} - 1} - \hat{A} \right\} \right.\\
         &\left. - \frac{\hat{A} - 1}{\hat{A}^2} \eta^2 s_{12}^2 c_{12}^2 c_{23}^2 \left\{ \cos{(\frac{\Delta m^2_{21}L}{2E})} - 1 \right\} 
         -\frac{s_{13}^2 s_{23}^2}{\hat{A} - 1} \left\{ \cos{(\frac{\Delta m^2_{31}L}{2E})} - 1 \right\} \right]~.
    \end{aligned}
\end{equation}

\begin{figure*}[!htbp]
\includegraphics[width=0.32\linewidth]{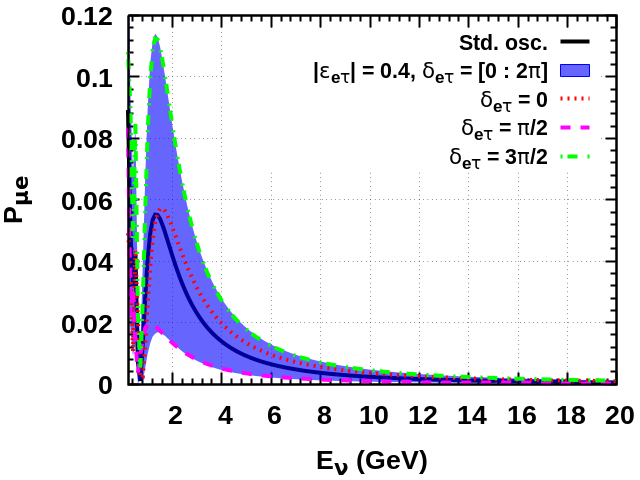}
\includegraphics[width=0.32\linewidth]{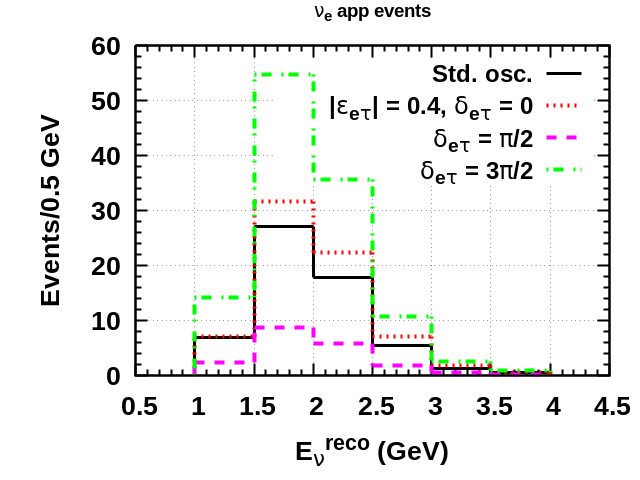}
\includegraphics[width=0.32\linewidth]{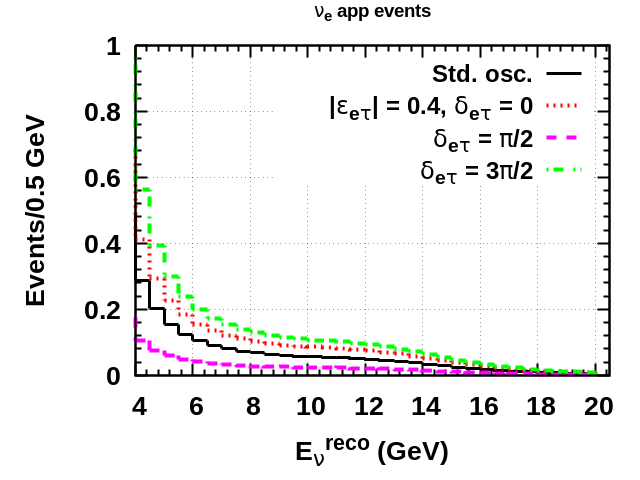}
\includegraphics[width=0.32\linewidth]{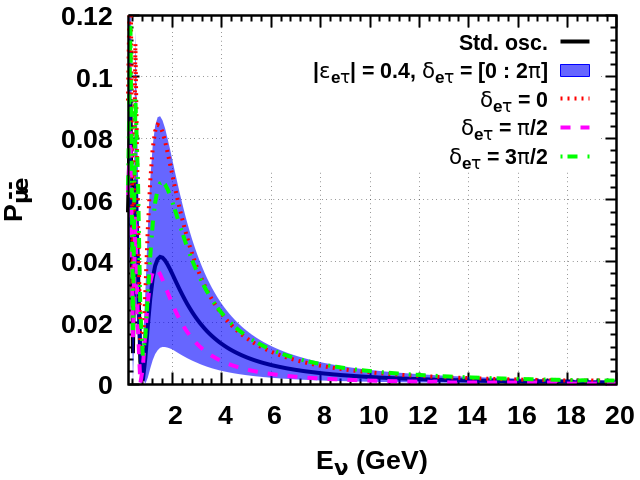}
\includegraphics[width=0.32\linewidth]{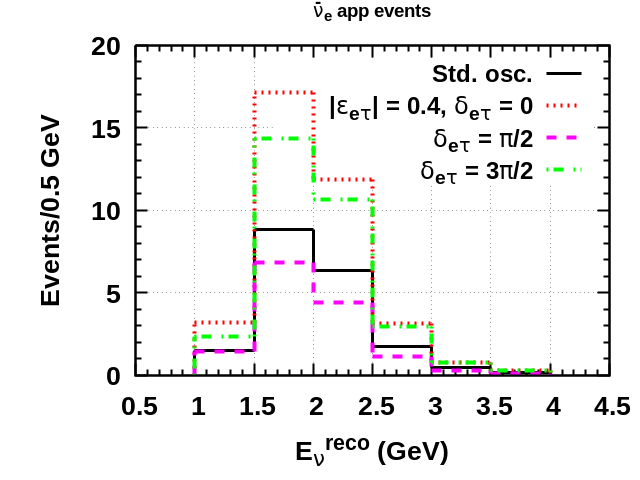}
\includegraphics[width=0.32\linewidth]{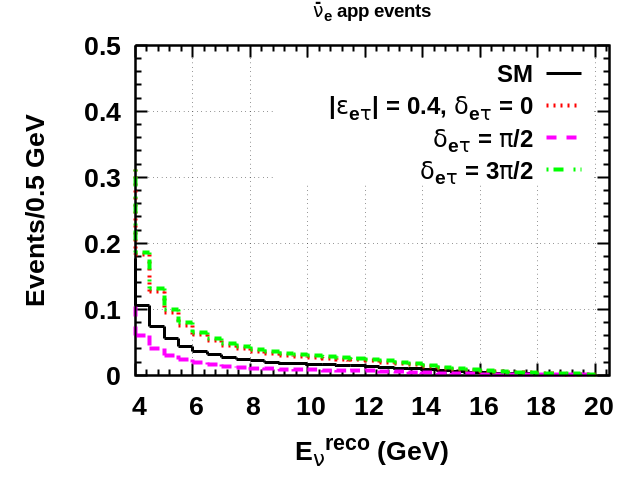}
\caption{Appearance probability versus neutrino energy and event rates versus reconstructed energy for $\nu_e$ ($\bar{\nu}_e$) in the upper (lower) row. We consider $|\epsilon_{e\tau}| = 0.4$ and corresponding phases to show modified probability and event rates in the presence of non-standard interaction.}
\label{fig:prob-event-eps_etau}
\end{figure*}
Here we use the following abbreviation: $\hat{A} \equiv \frac{A}{\Delta m^2_{31}},~s_{ij} \equiv \sin{\theta_{ij}}$ and $c_{ij} = \cos{\theta_{ij}}$~. We can obtain the probabilities corresponding to $\bar{\nu}_\mu \rightarrow \bar{\nu}_e$ channel replacing $A \rightarrow -A,~ \delta_{CP} \rightarrow -\delta_{CP},~ \delta_{e\tau} \rightarrow -\delta_{e\tau}$.
We present the numerical appearance probability $P_{\mu e}~(P_{\bar{\mu}\bar{e}})$ versus neutrino energy $E_\nu$ and simulated event rates versus reconstructed energy $E_\nu^{reco}$ in fig.~\ref{fig:prob-event-eps_etau} using eq.~(\ref{eq:Prob}). Upper and lower rows correspond to $\nu_\mu \rightarrow \nu_e$ and $\bar{\nu}_\mu \rightarrow \bar{\nu}_e$ channels. The plots in the left panel represent probability, while other plots illustrate the event rates. The middle and right panels correspond to events per bin in the energy ranges $1<E_\nu<4$ GeV and $4<E_\nu<20$ GeV, respectively. 

In each of the probability plots (left panel), the black curve represents standard oscillation probability, whereas the blue band refers to oscillation probability embodying the effect of $|\epsilon_{e\tau}| = 0.4$ with the variable non-standard phase $\delta_{e\tau} = [0:2\pi]$. The red dotted, magenta dashed and green dot-dashed curves include $|\epsilon_{e\tau}| = 0.4$ with different phases $\delta_{e\tau} = 0$, $\delta_{e\tau} = \pi/2$ and $\delta_{e\tau} = 3\pi/2$ respectively. For $|\epsilon_{e\tau}| = 0.4$ and $\delta_{e\tau} = 0, 3\pi/2$, $P_{\mu e}$ is higher than the standard oscillation probability as can be seen from red dotted and green dot-dashed lines, respectively. On the other hand, $|\epsilon_{e\tau}| = 0.4$, $\delta_{e\tau} = \pi/2$ (magenta) shows lower probability w.r.t standard oscillation. In the case of $P_{\bar{\mu}\bar{e}}$ the calculated probability is higher than the standard probability for $\delta_{e\tau} = 0, 3\pi/2$ and lower than std.~osc.~for $\delta_{e\tau} = \pi/2$ as can be seen from the red, green and magenta curves respectively.
The standard oscillation curve is degenerate with NSI probability for a combination of $|\epsilon_{e\tau}|$ and $\delta_{e\tau}$ throughout the energy range. We notice the overlap between the blue band and solid black curve that indicates degeneracy between NSI ($|\epsilon_{e\tau}|$, $\delta_{e\tau}$) and SM case.

In the middle and right panels we depict events spectra corresponding to $\nu_e$ (top row) and $\bar{\nu}_e$ (bottom row) appearance channels. We show event rates related to energy ranges $1<E_\nu<4$ GeV (middle) and $4<E_\nu<20$ GeV (right). The legends are mentioned in the respective plots. 


\subsection{Allowed regions in \texorpdfstring{$\epsilon_{e\tau}$}~~and \texorpdfstring{$\delta_{CP (e\tau)}$}~~parameter space}

\begin{figure*}[!htbp]
\includegraphics[width=0.45\linewidth]{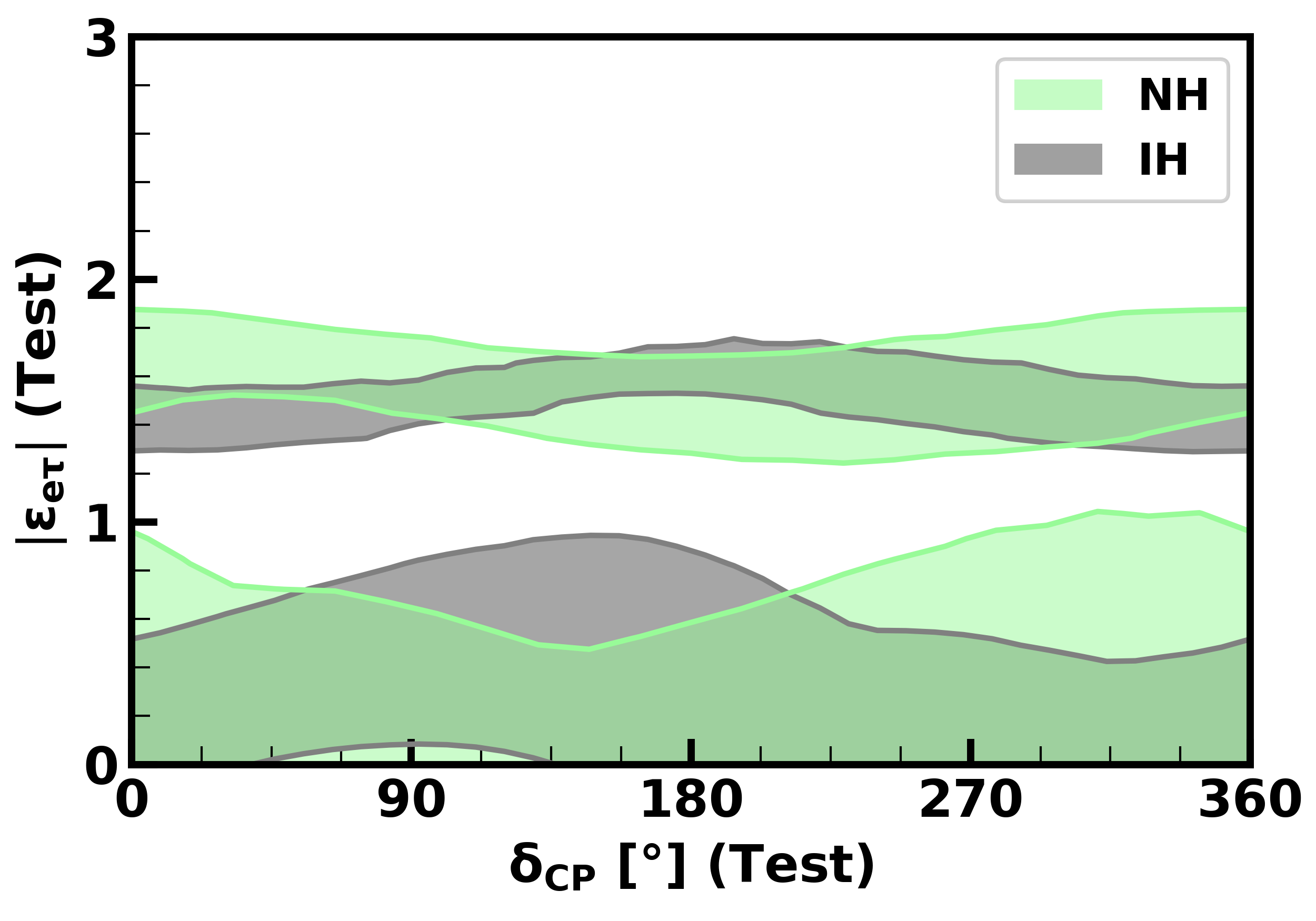}
\includegraphics[width=0.45\linewidth]{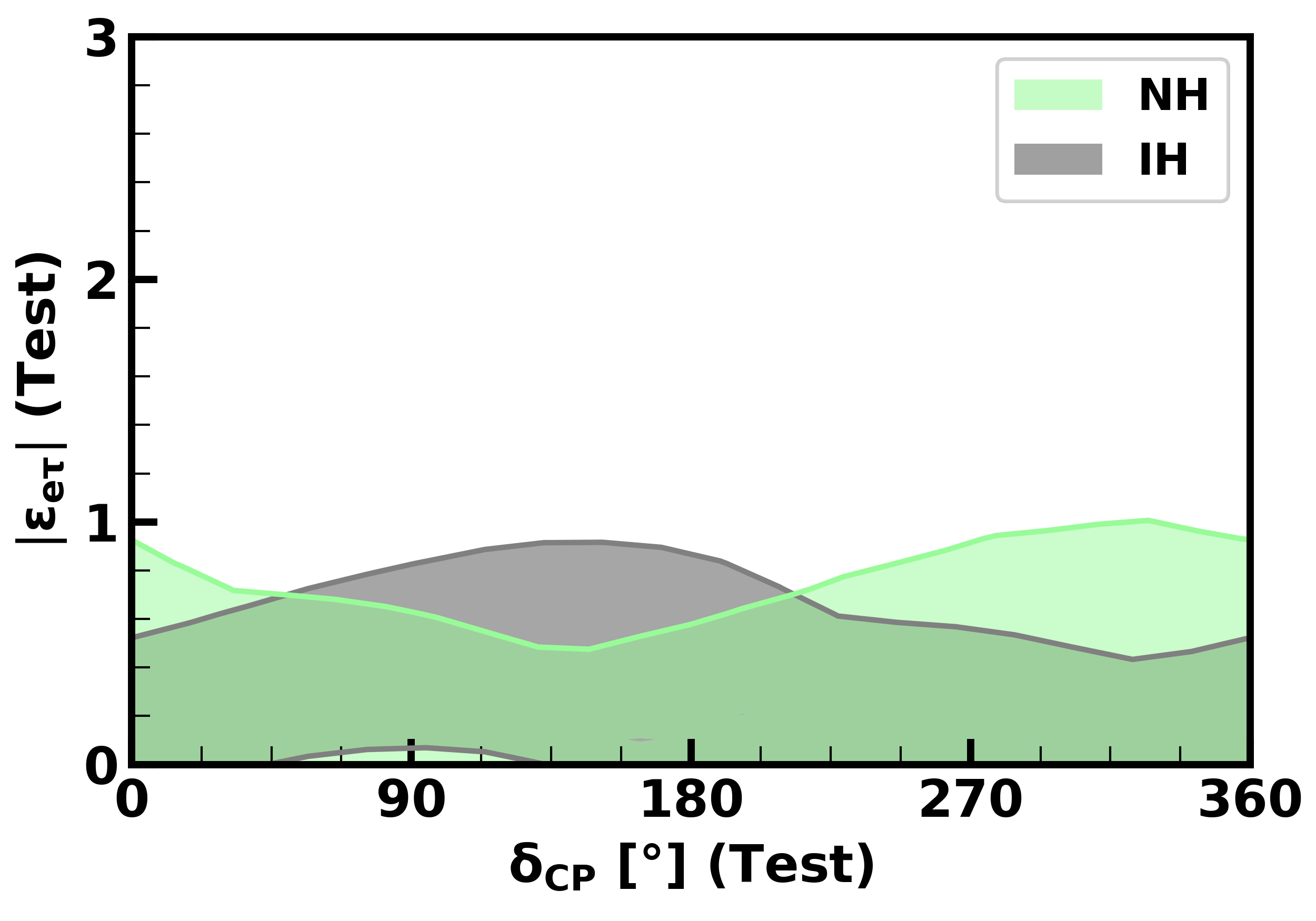}
\includegraphics[width=0.45\linewidth]{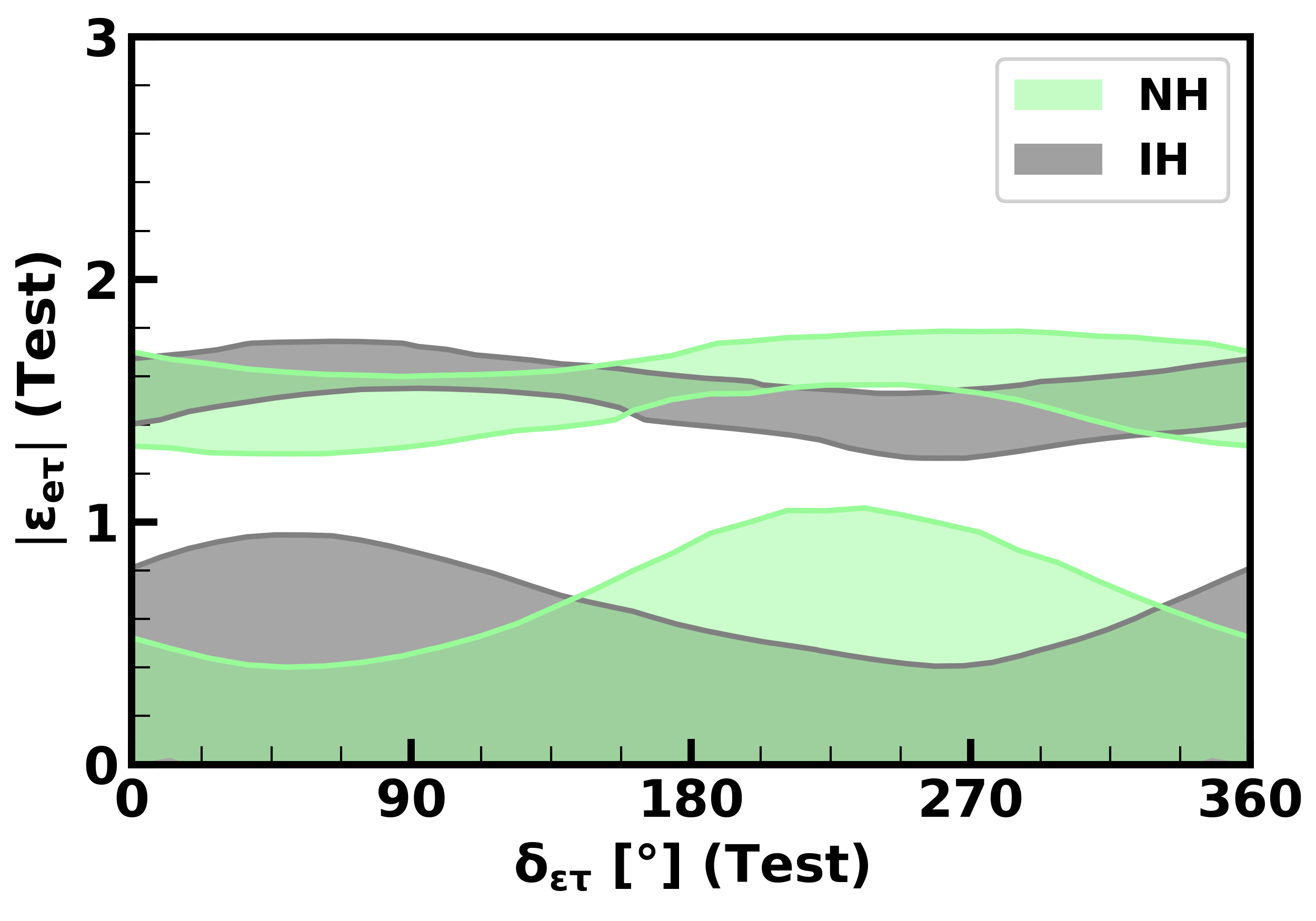}
\includegraphics[width=0.45\linewidth]{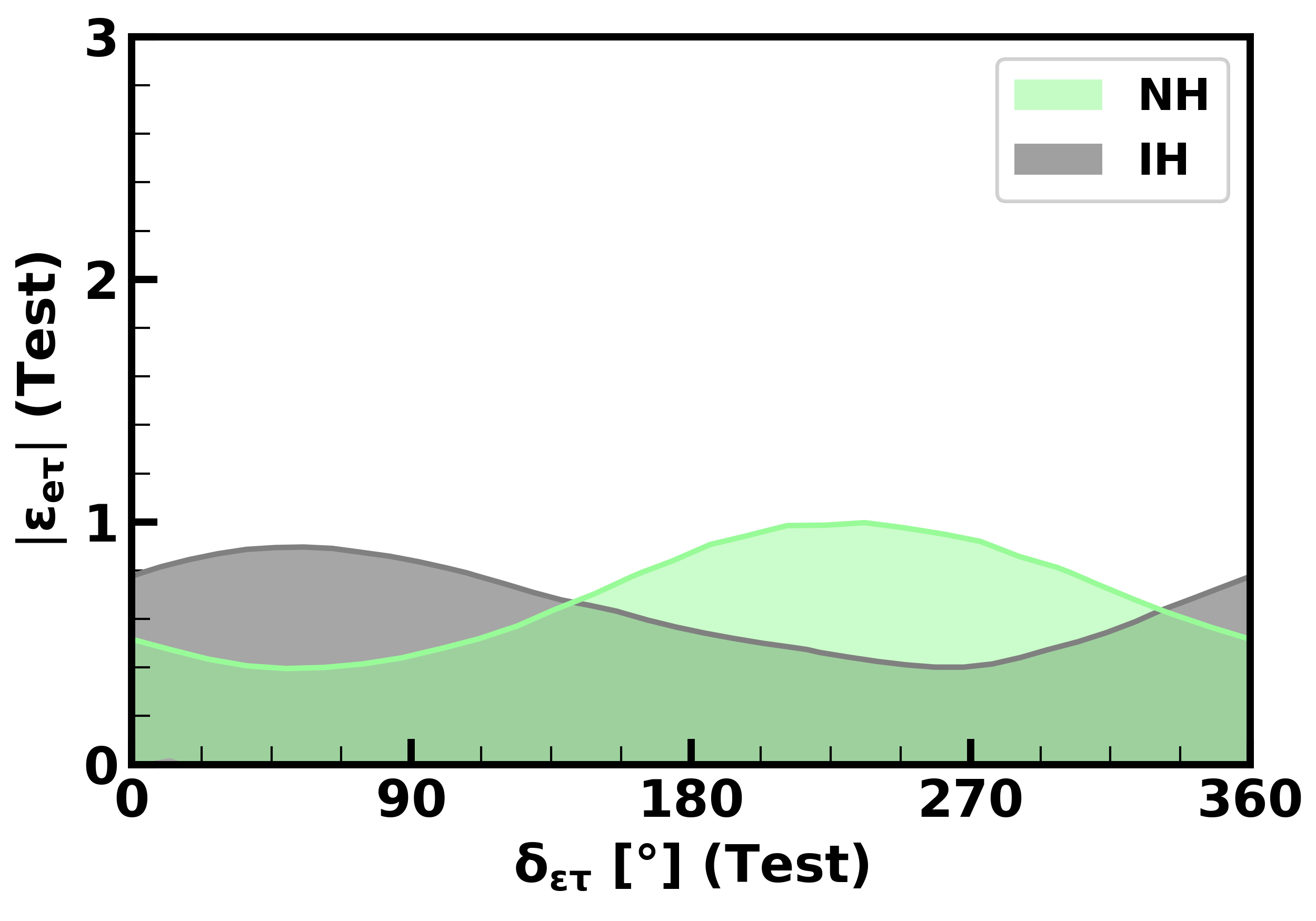}
\caption{$\epsilon_{e\tau}$ (test) vs $\delta_{CP}$ (test) in the upper row and $\epsilon_{e\tau}$ (test) vs $\delta_{e\tau}$ (test) in the lower row. In the left panel 1 - 4 GeV and in the right panel 1 - 20 GeV. Marginalized over $\theta_{13}$, $\theta_{23}$, $\Delta m^2_{31}$, $\delta_{e\tau}$ ($\delta_{CP}$) in upper row (lower row).}
\label{fig:eps_etau-vs-delcp_del_etau}
\end{figure*}

In fig.~\ref{fig:eps_etau-vs-delcp_del_etau} we show the allowed contours with $95\%$ ($1.96 \sigma$) CL in the $|\epsilon_{e\tau}|-\delta_{CP}$ plane (top row) and $|\epsilon_{e\tau}|-\delta_{e\tau}$ plane (bottom row). Plots shown in the left and right panels considering energy analysis energy window $1<E_\nu<4$ GeV and $1<E_\nu<20$ GeV, respectively. We perform the analysis using standard model scenario in the simulated data and only non-zero $\epsilon_{e\tau}$ in the theory. We compute $\Delta \chi^2$ as per the definition,
\begin{equation}
    \Delta \chi^2 = \chi^2 \left(\epsilon_{e\tau} [true] = 0, \epsilon_{e\tau} [test] \neq 0 \right) ,
    \label{eq:chi-sq-eps_etau}
\end{equation}
with true $\delta_{CP}$ fixed at $0.82\pi$ and marginalize over $\theta_{13},~\theta_{23},~\Delta m^2_{31}$ and $\delta_{e\tau}$. 
Similarly, we calculate $\Delta \chi^2$ corresponding to $\epsilon_{e\tau}$[test] versus $\delta_{e\tau}$[test] contour using eq.~\ref{eq:chi-sq-eps_etau}. Here, we marginalize over $\theta_{13},~\theta_{23},~\Delta m^2_{31}$ and $\delta_{CP}$. We vary standard and non-standard CP phase in the full range $0-360\degree$ and show regions with $95\%$ CL for both the normal hierarchy (NH, green regions) and inverted hierarchy (IH, gray region). The unmentioned oscillation parameters are kept fixed to their true values.

\begin{figure*}[!htbp]
\includegraphics[width=0.45\linewidth]{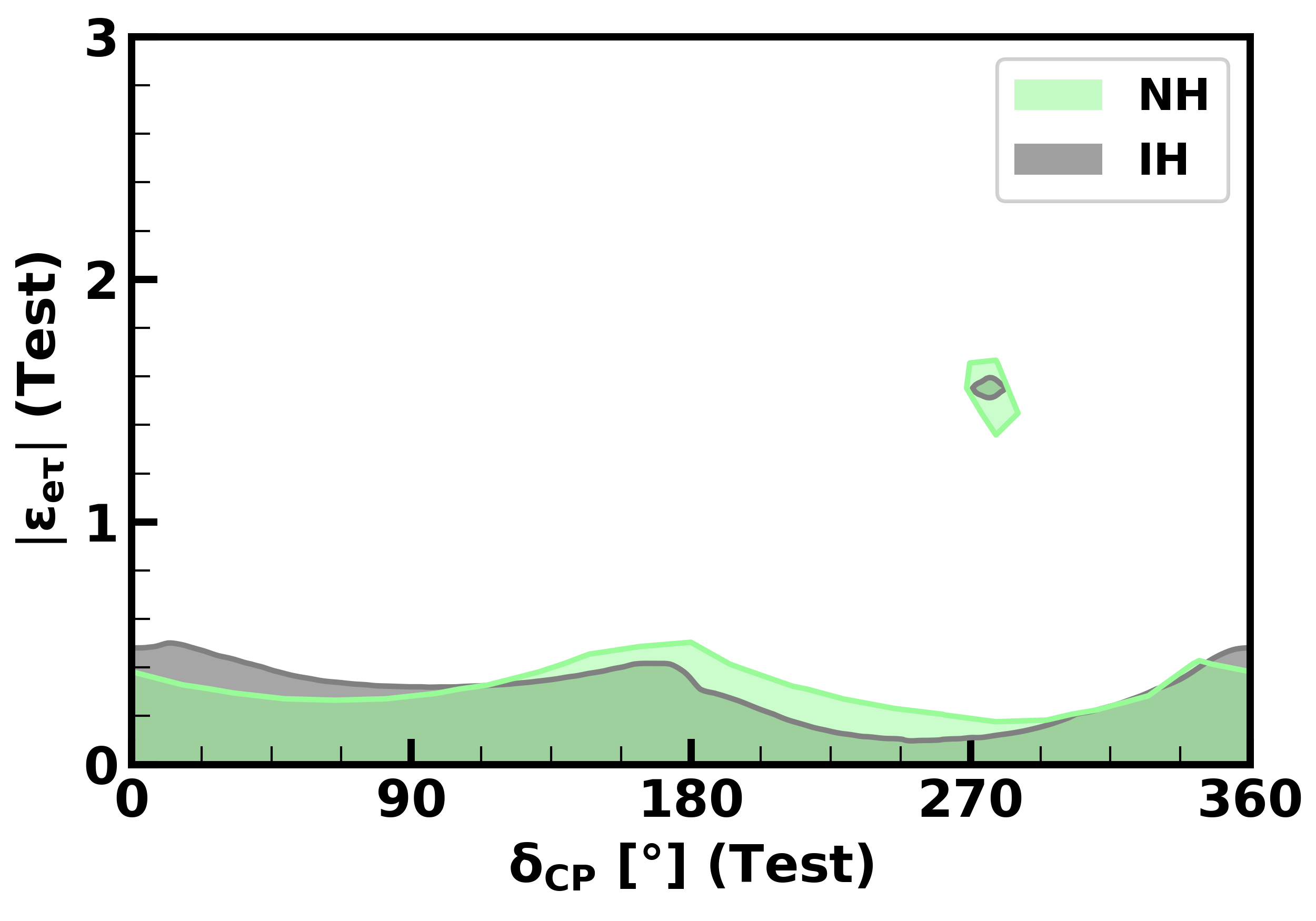}
\includegraphics[width=0.45\linewidth]{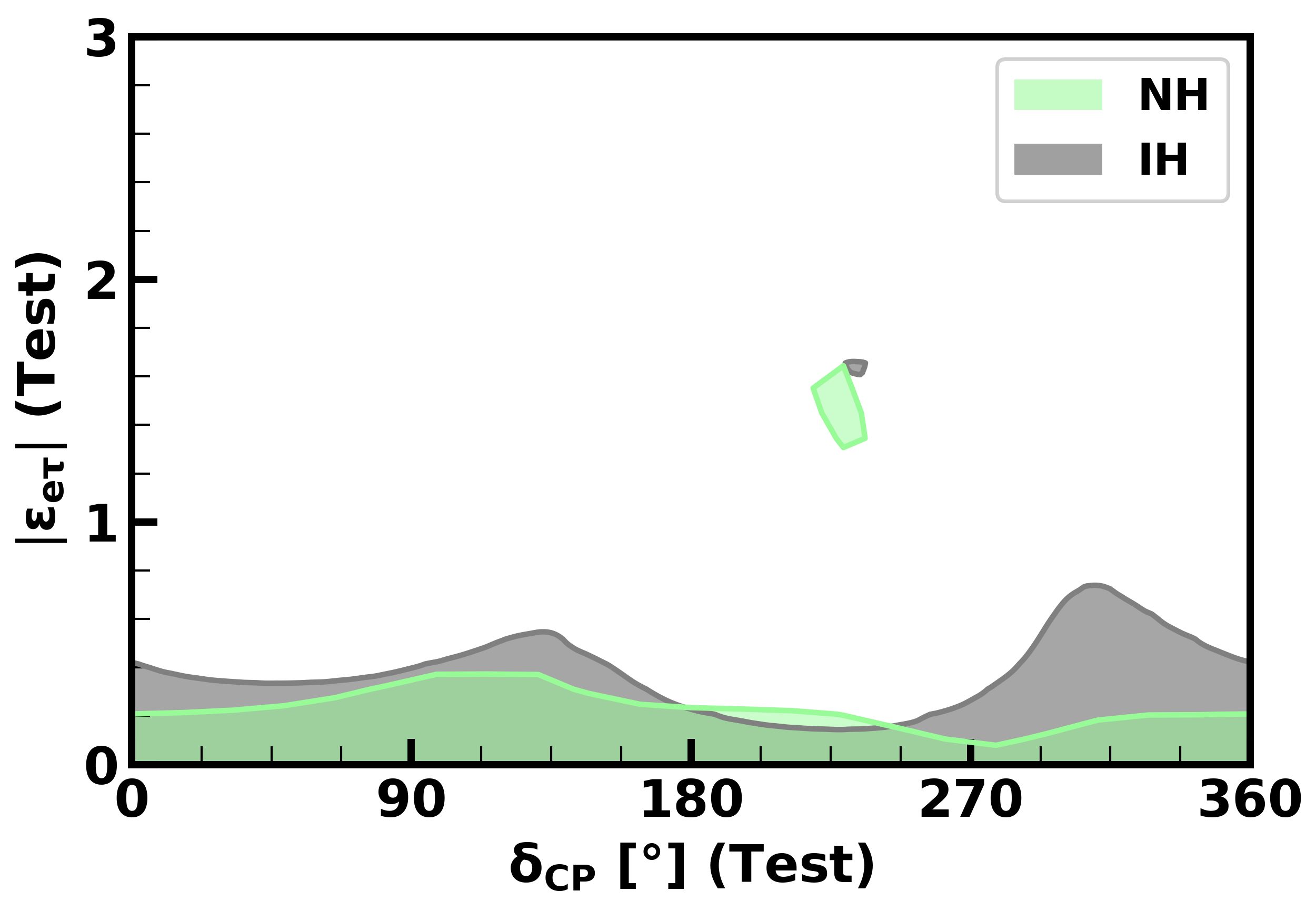}
\caption{$\epsilon_{e \tau}$ vs $\delta_{CP}$ for fixed $\delta_{e\tau} = 0$ in left and $\delta_{e\tau} = 45\degree$ in right.}
\label{fig:eps_etau-vs-delcp_fix_del_etau}
\end{figure*}

\begin{figure*}[!htbp]
\includegraphics[width=0.325\linewidth]{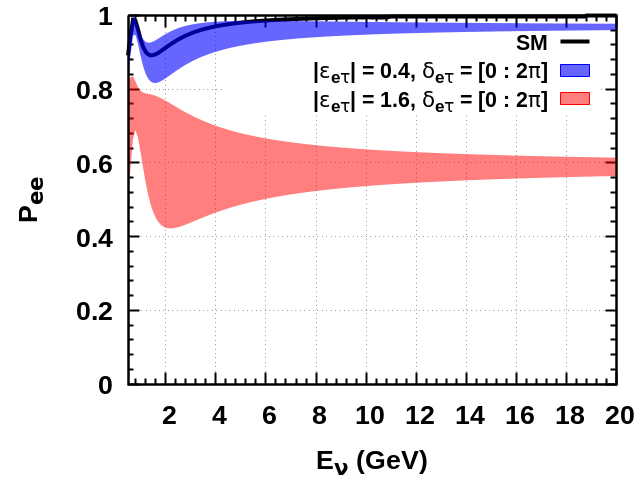}
\includegraphics[width=0.325\linewidth]{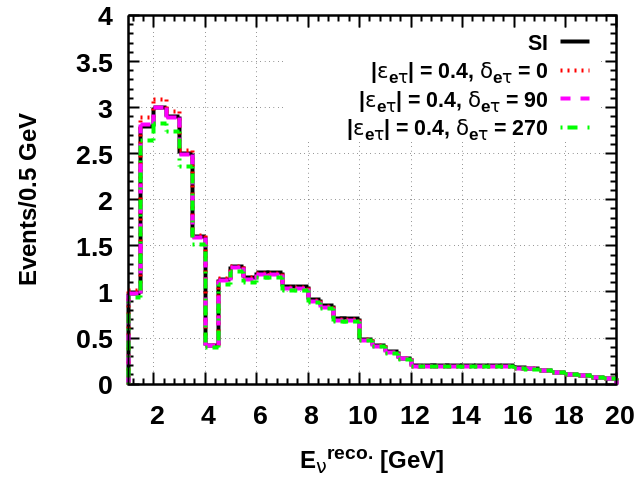}
\includegraphics[width=0.325\linewidth]{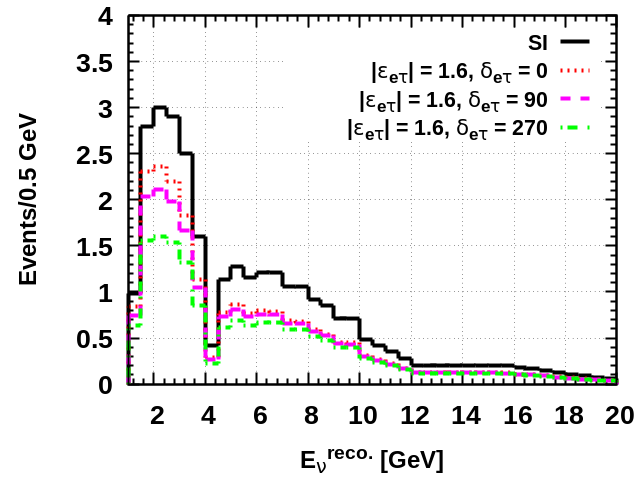}
\caption{$P_{ee}$ vs energy (left) and beam background events per bin with reconstructed energy (middle and right). In the middle $|\epsilon_{e\tau}| = 0.4$ and in the right $|\epsilon_{e\tau}| = 1.6$. We show the plots for different $\delta_{e\tau} = 0, 90\degree, 270\degree$ in both the figures.}
\label{fig:bkg_events-vs-energy}
\end{figure*}

In the top-left plot of Fig.~\ref{fig:eps_etau-vs-delcp_del_etau} we can see from the green bands (NH) that for each value of $\delta_{CP}$, two bands span the allowed range in $\epsilon_{e \tau}$ parameter space indicating a degeneracy of the form $(\epsilon, \delta) = (\epsilon', \delta')$. Similar behavior is observed for the inverted hierarchy case and can be seen from the gray curve. These results agree with the results in ref.~\cite{Kleykamp:2022dli}. 
After including high-energy simulated events, plots shown in the right panel, we see the degeneracy in the $|\epsilon_{e\tau}|$ is removed for whole range of standard CP-phase $\delta_{CP}$ and corresponding non-standard phase $\delta_{e\tau}$. These observations are true for both the normal and inverted mass hierarchies. We investigate this further in the following.

Firstly to understand the degeneracy in the left panels of fig~\ref{fig:eps_etau-vs-delcp_del_etau}, we plot $\delta_{CP}$ vs $\epsilon_{e \tau}$ for fixed $\delta_{e\tau}$, $\delta_{e\tau} = 0$ in the left panel and $\delta_{e\tau} = 45\degree$ in the right panel of fig.~\ref{fig:eps_etau-vs-delcp_fix_del_etau}. From the left plot, we can see that there is a small degenerate region around $\epsilon_{e\tau} \sim 1.6$ corresponding to the $\delta_{CP} \sim 270\degree$ for both NH and IH bands. Similarly, in the right plot there is a degenerate region for  $\epsilon_{e\tau} \sim 1.6$ around $\delta_{CP} \sim 225\degree$ i.e. for $\delta_{CP} \sim (270 - 45)$. This can be generalized to the form $\delta_{CP} = 270 - \delta_{e\tau}$ or $3 \pi/2$. This can be attributed to a degeneracy in the $\nu_e$ and $\bar{\nu}_e$ spectrum between SM and NSI $(\epsilon_{e\tau} \sim 1.6$,  $\delta_{CP} = 270 - \delta_{e\tau}$). This agrees with the conclusions drawn in NOvA collaboration paper~\cite{NOvA:2024lti}.

Further, we investigate the results obtained in the right panel of fig~\ref{fig:eps_etau-vs-delcp_del_etau}, where the degenerate bands around $\epsilon_{e\tau} \sim 1.6$ disappear. Here we are adding the high energy simulated events from $4 < E_\nu < 20$ GeV. Note that this high energy tail has both signal and background events contributing to the $\chi^2$ as can be seen from fig.~\ref{fig:prob-event-eps_etau} (signal events) and fig.~\ref{fig:bkg_events-vs-energy} (background events). Clearly, in comparison, the beam background is dominant compared to the signal events in this energy window. This can be verified from the $(4 - 20)$ GeV events given in the table~\ref{table:signal_bkg_events}. To investigate the role of background in removing the degeneracy, we further look into the relevant oscillation channel $P_{ee}$. Therefore, we illustrate the probability vs $E_\nu$ (left panel) .and events vs $E_\nu$ (assuming $\epsilon_{e\tau} = 0.4$ and $\epsilon_{e\tau} = 1.6$ in the middle and right panels) for the beam background channel $P(\nu_e \rightarrow \nu_e)$ in fig.~\ref{fig:bkg_events-vs-energy}.
In all three plots, the relevant NSI parameters are as listed in the legends. In the left plot, the blue band (corresponding to $|\epsilon| = 0.4$ and $\delta_{e\tau} \in [0:2\pi]$) has a significant overlap with the SM curve (black curve) and, on the contrary, the red band (corresponding to $|\epsilon| = 1.6$ and $\delta_{e\tau} \in [0:2\pi]$) is well separated from the SM curve, indicating that the degeneracy between SM and NSI $(\epsilon_{e\tau} \sim 1.6$,  $\delta_{CP} = 270 - \delta_{e\tau}$) can be broken when these events are considered in the analysis~\footnote{We verified that in the case of signal probability $P_{\mu e}$ both the blue and red bands overlap with the SM curve $4 < E_\nu < 20$ GeV.}. In the middle plot, the curves indicating different values of $\delta_{e \tau}$, $\epsilon_{e\tau} = 0.4$ are close to the SM curve (black) showing very less sensitivity to distinguish between SM and NSI. However, in the right panel the reasonable gap between the SM and NSI curves leads to the elimination of degeneracy in Fig.~\ref{fig:eps_etau-vs-delcp_del_etau} (right panels). 

Furthermore, the $\chi^2$ analysis in Fig.~\ref{fig:eps_etau-delcp_deletau-no_sig-5-20GeV} of the appendix~\ref{app:eps_etau_delcp-no_sig} validates the role of beam background from high energy tail in removing the degeneracy $\epsilon_{e\tau}\sim 1.6$. We show $95\%$ CL region in the $|\epsilon_{e\tau}|$ vs $\delta_{CP}$ ($\delta_{e\tau}$) plane in the left (right) plot considering only background (no signal) in the $4<E_\nu<20$ GeV energy window. We observe a narrow degeneracy band around $|\epsilon_{e\tau}| \sim 1.6$ in the left plot for the $\delta_{CP}\in [190\degree : 270\degree]$ in the case of NH. Other than that the degeneracy around $|\epsilon_{e\tau}| \sim 1.6$ is eliminated in both the left and right plots throughout the $\delta_{CP}$ and $\delta_{e\tau}$ range in case of both NH and IH. This observation infers that the background events from high energy tail contribute to eliminate the degeneracy, except the narrow band in the left plot for the $\delta_{CP}\in [190\degree : 270\degree]$ in the case of NH.

    \begin{table}[!htb]
      \setlength{\tabcolsep}{5pt}
  \begin{tabular}{|l|l|l|l|l|l|l|l|}
    \hline
    \multicolumn{8}{|c|}{Signal events} \\
    \hline
    \multicolumn{2}{|c|}{\multirow{3}{*}{Hypothesis}} &
      \multicolumn{3}{c|}{$\nu_\mu \rightarrow \nu_e$} &
      \multicolumn{3}{c|}{$\bar{\nu}_\mu \rightarrow \bar{\nu}_e$} \\
      \cline{3-8}
      \multicolumn{2}{|c|}{\multirow{2}{*}{}} & 1 - 4 & 1 - 20 & 4 - 20 & 1 - 4 & 1 - 20 & 4 - 20\\
      \multicolumn{2}{|c|}{\multirow{2}{*}{}} & GeV & GeV & GeV & GeV & GeV & GeV\\
      \hline
      \multicolumn{2}{|c|}{SM} & 59.0 & 61.0 & 2.0 & 19.0 & 19.6 & 0.6\\
      \hline
      {\multirow{3}{*}{$|\epsilon_{e\tau}|=0.4$}} & $\delta_{e\tau} = 0$ & 70.0 & 73 & 3.0 & 36.3 & 37.4 & 1.1\\
      \cline{2-8}
      & $\delta_{e\tau} = \pi/2$ & 19.0 & 20.0 & 1.0 & 14.1 & 14.4 & 0.3\\
      \cline{2-8}
      & $\delta_{e\tau} = 3\pi/2$ & 118.3 & 122.3 & 4.0 & 31.3 & 32.4 & 1.1\\
      \hline
      \multicolumn{8}{|c|}{Beam background ($\nu_e + \bar{\nu}_e$)} \\
      \hline
      \multicolumn{2}{|c|}{SM} & 13.7 & 29.6 & 15.9 & 7.6 & 16.4 & 8.8\\
      \hline
      {\multirow{3}{*}{$|\epsilon_{e\tau}|=0.4$}} & $\delta_{e\tau} = 0$ & 14.1 & 29.7 & 15.6 & 7.8 & 16.5 & 8.7\\
      \cline{2-8}
      & $\delta_{e\tau} = \pi/2$ & 13.7 & 29.3 & 15.6 & 7.6 & 16.2 & 8.6\\
      \cline{2-8}
      & $\delta_{e\tau} = 3\pi/2$ & 13.0 & 28.2 & 15.2 & 7.2 & 15.6 & 8.4\\
      \hline
      {\multirow{3}{*}{$|\epsilon_{e\tau}|=1.6$}} & $\delta_{e\tau} = 0$ & 10.6 & 20.9 & 10.3 & 5.9 & 11.6 & 5.7\\
      \cline{2-8}
      & $\delta_{e\tau} = \pi/2$ & 9.5 & 19.4 & 9.9 & 5.3 & 10.7 & 5.4\\
      \cline{2-8}
      & $\delta_{e\tau} = 3\pi/2$ & 7.5 & 16.3 & 8.8 & 4.1 & 9.0 & 4.9\\
    
    \hline
  \end{tabular}
  \caption{Signal and background events at HE considering $\epsilon_{e\tau} \neq 0$~.}
  \label{table:signal_bkg_events}
\end{table}

\subsection{Oscillation probabilities and event rates in the presence of \texorpdfstring{$\epsilon_{e\mu}$}~}

In this subsection, we illustrate the appearance probabilities and event rates w.r.t neutrino energy assuming non-zero $\epsilon_{e\mu}$. The approximate probability corresponding to $\nu_\mu \rightarrow \nu_e$ channel considering non-zero $\epsilon_{e\mu}$ is 
\begin{equation}\label{eq:app-prob-approx-eps_emu}
    \begin{aligned}
        P_{\mu e} =& 2 \left[ \frac{1}{\hat{A} - 1} \left\{ \frac{\eta s_{12} c_{12} s_{13} s_{23} c_{23}}{\hat{A}} + \eta s_{12} c_{12} s_{23}^2 c_{23} \epsilon_{e\mu} + s_{13} s_{23} c_{23}^2 \epsilon_{e\mu} + \hat{A} s_{23}^2 c_{23}^2 \epsilon_{e\mu}^2 \right\} \right.\\
        &\left. \times \left\{ - \cos{(\frac{\Delta m^2_{21}L}{2E})} - \cos{(\frac{\Delta m^2_{31}L}{2E})} + \cos{(\frac{\Delta m^2_{32}L}{2E})} + 1 \right\} \right.\\
        &\left. - \frac{1}{\hat{A}^2} \left\{ \eta^2 s_{12}^2 c_{12}^2 c_{23}^2 + 2\eta \hat{A} s_{12} c_{12} c_{23}^3 \epsilon_{e\mu} + \hat{A}^2 c_{23}^4 \epsilon_{e\mu}^2 \right\} \left\{ \cos{(\frac{\Delta m^2_{21}L}{2E})} - 1 \right\} \right.\\
        &\left. - \frac{1}{(\hat{A} - 1)^2} \left\{ s_{13}^2 s_{23}^2 + 2\hat{A} s_{13} s_{23}^3 \epsilon_{e\mu} + \hat{A}^2 s_{23}^4 \epsilon_{e\mu}^2 \right\} \left\{ \cos{(\frac{\Delta m^2_{31}L}{2E})} - 1 \right\} \right]~.
    \end{aligned}
\end{equation} 

\begin{figure*}[!htbp]
\includegraphics[width=0.32\linewidth]{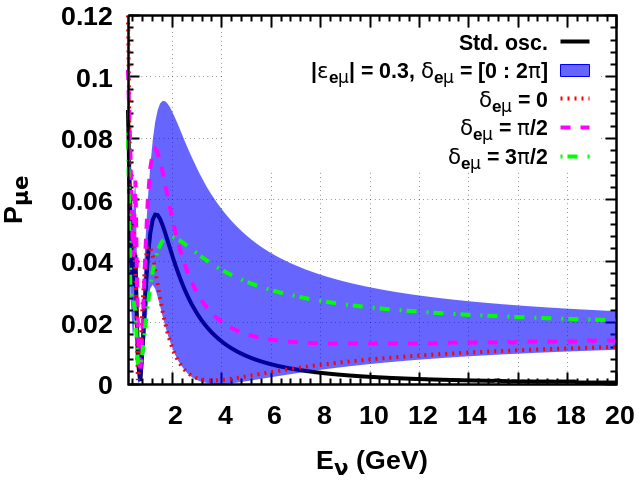}
\includegraphics[width=0.32\linewidth]{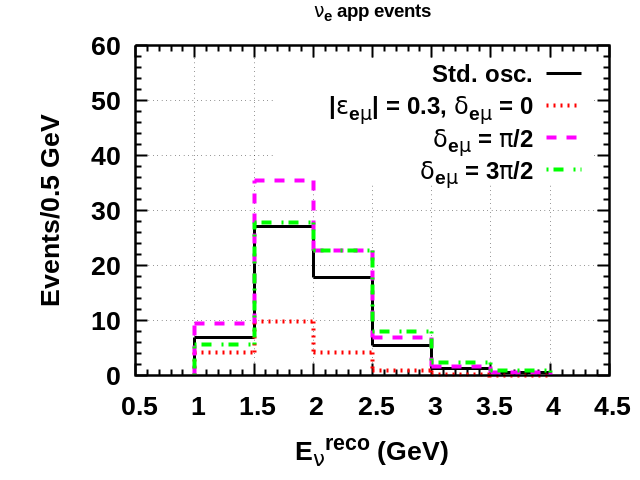}
\includegraphics[width=0.32\linewidth]{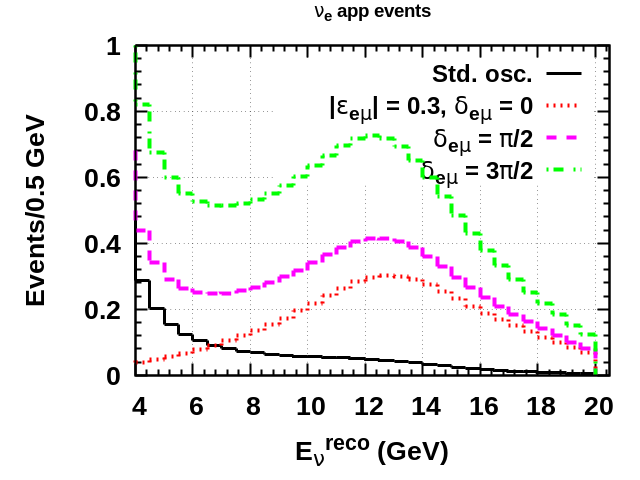}
\includegraphics[width=0.32\linewidth]{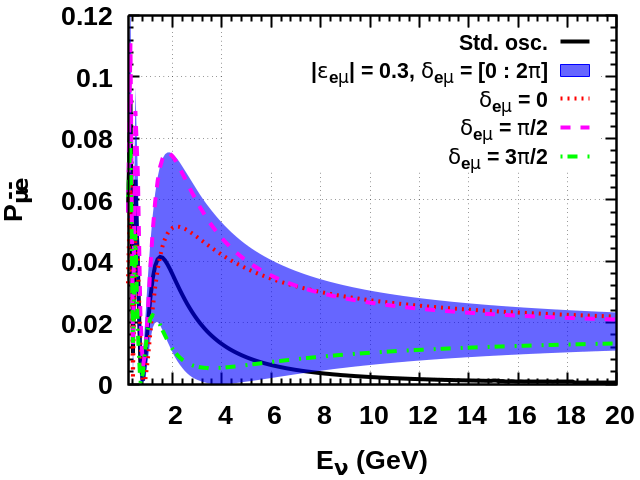}
\includegraphics[width=0.32\linewidth]{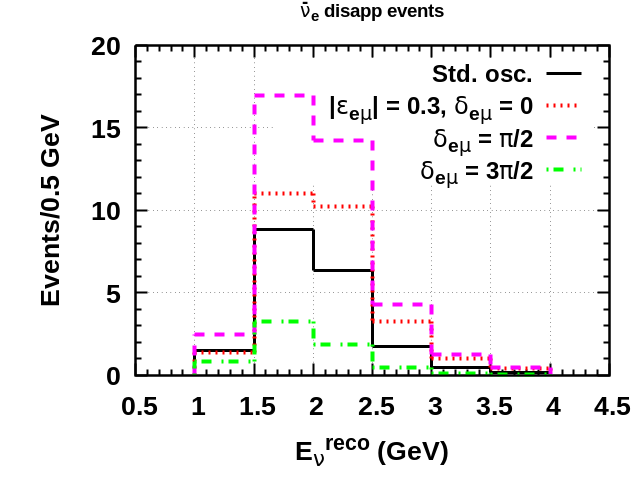}
\includegraphics[width=0.32\linewidth]{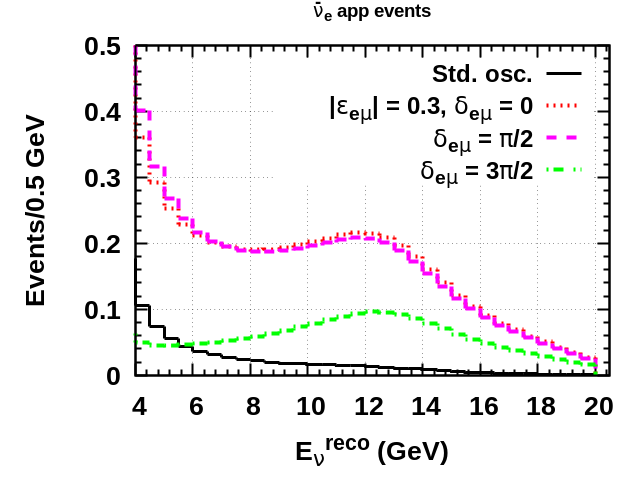}
\caption{Appearance probability versus neutrino energy and event rates versus reconstructed energy for $\nu_e$ ($\bar{\nu}_e$) in the upper (lower) row. We consider $|\epsilon_{e\mu}| = 0.3$ and corresponding phases to show modified probability and event rates in the presence of non-standard interaction.}
\label{fig:prob-event-eps_emu}
\end{figure*}

In the left panel of  fig.~\ref{fig:prob-event-eps_emu} we show $\nu_e$ (upper) and $\bar{\nu}_e$ (lower) numerical appearance probabilities using eq.~(\ref{eq:Prob}). In each of the probability plots we draw five curves; the black curve represents standard oscillation probability, the blue band depicts the resultant probability for $|\epsilon_{e\mu}| = 0.3$ and $\delta_{e\mu} \in [0:2\pi]$, the red dotted, magenta dashed and green dot-dashed curves stand for nonstandard phases $\delta_{e\mu} = 0$, $\delta_{e\mu} = \pi/2$ and $\delta_{e\mu} = 3\pi/2$ respectively. In all the plots we assume NH as the true mass hierarchy.

From the left panel of fig.~\ref{fig:prob-event-eps_emu} the blue band corresponding to non-zero $|\epsilon_{e\mu}| = 0.3$ and $\delta_{e\mu} \in [0:2\pi]$ spans a wide range of probability values both greater and lesser than the standard oscillation probability (black curve). Whereas for $E_\nu>7$ GeV, we do not observe degeneracy for the chosen parameters; rather, we see the increase in probabilities w.r.t standard oscillation. 
Additionally, we show the curves corresponding to CP conserving ($\delta_{e\mu} = 0$) and maximal CP violation ($\delta_{e\mu} = \pi/2,~3\pi/2$) using red, magenta and blue lines respectively.

In the middle and right panels we depict event rates as a function of reconstructed energy ranging from $1<E_\nu<4$ GeV and $4<E_\nu<20$ GeV respectively. The top and bottom panels correspond to $\nu_e$ and $\bar{\nu}_e$ appearance channels respectively. The relevant legends are listed within each plot. The $\nu_e$ and $\bar{\nu}_e$ appearance channels in the middle panel depict the event rates considered by the NOvA collaboration in ref.~\cite{NOvA:2024lti}. In the right most panel we show signal events corresponding to $E_\nu : 4-20$ GeV both for SM and assuming non-zero $\epsilon_{e\mu}$. The trends followed by events spectra in these panels are easily understandable from the corresponding probability plots of the left most panels.
\subsection{Allowed regions in \texorpdfstring{$\epsilon_{e\mu}$}~~and \texorpdfstring{$\delta_{CP (e\mu)}$}~~parameter space}

\begin{figure*}[!htbp]
\includegraphics[width=0.45\linewidth]{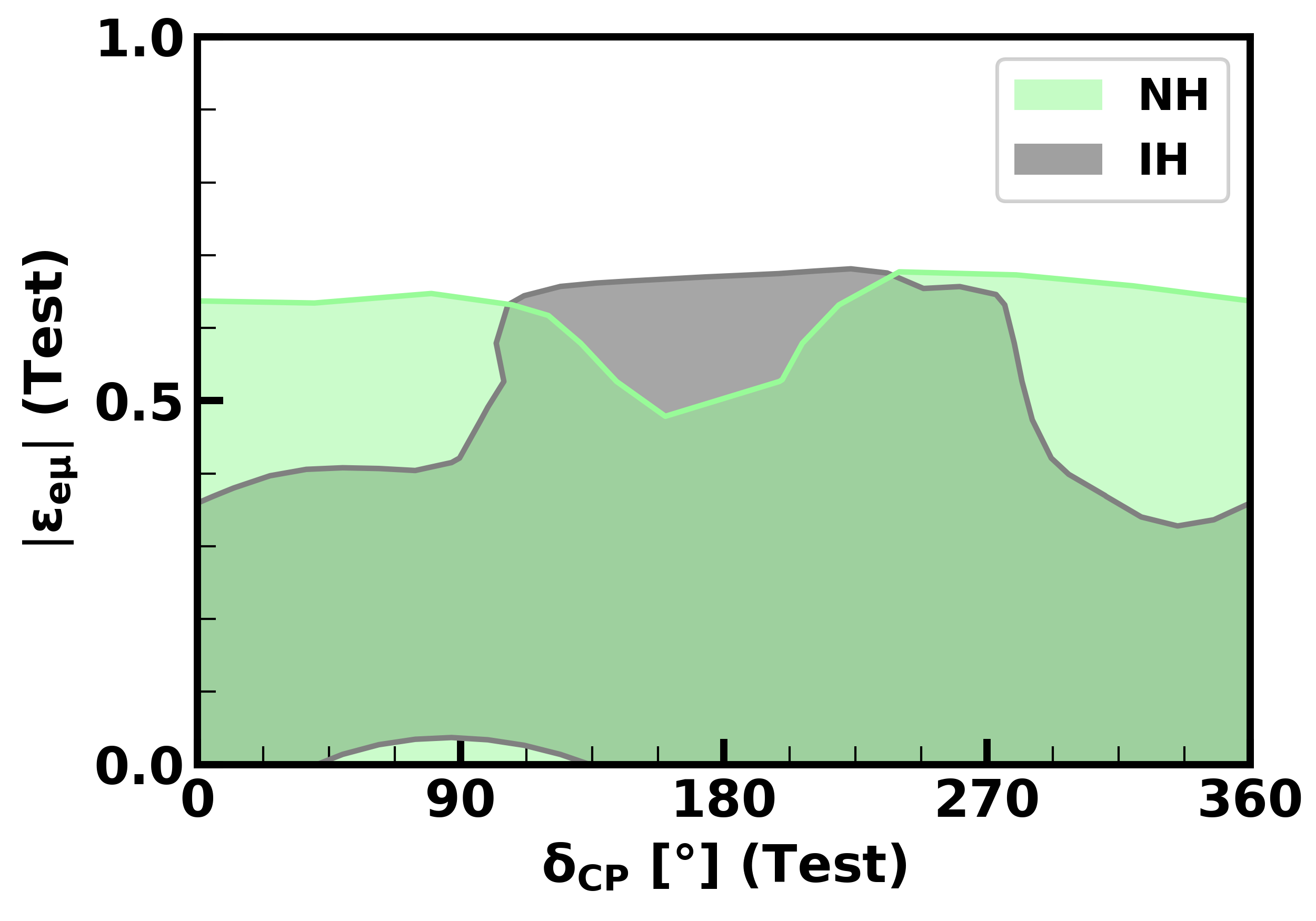}
\includegraphics[width=0.45\linewidth]{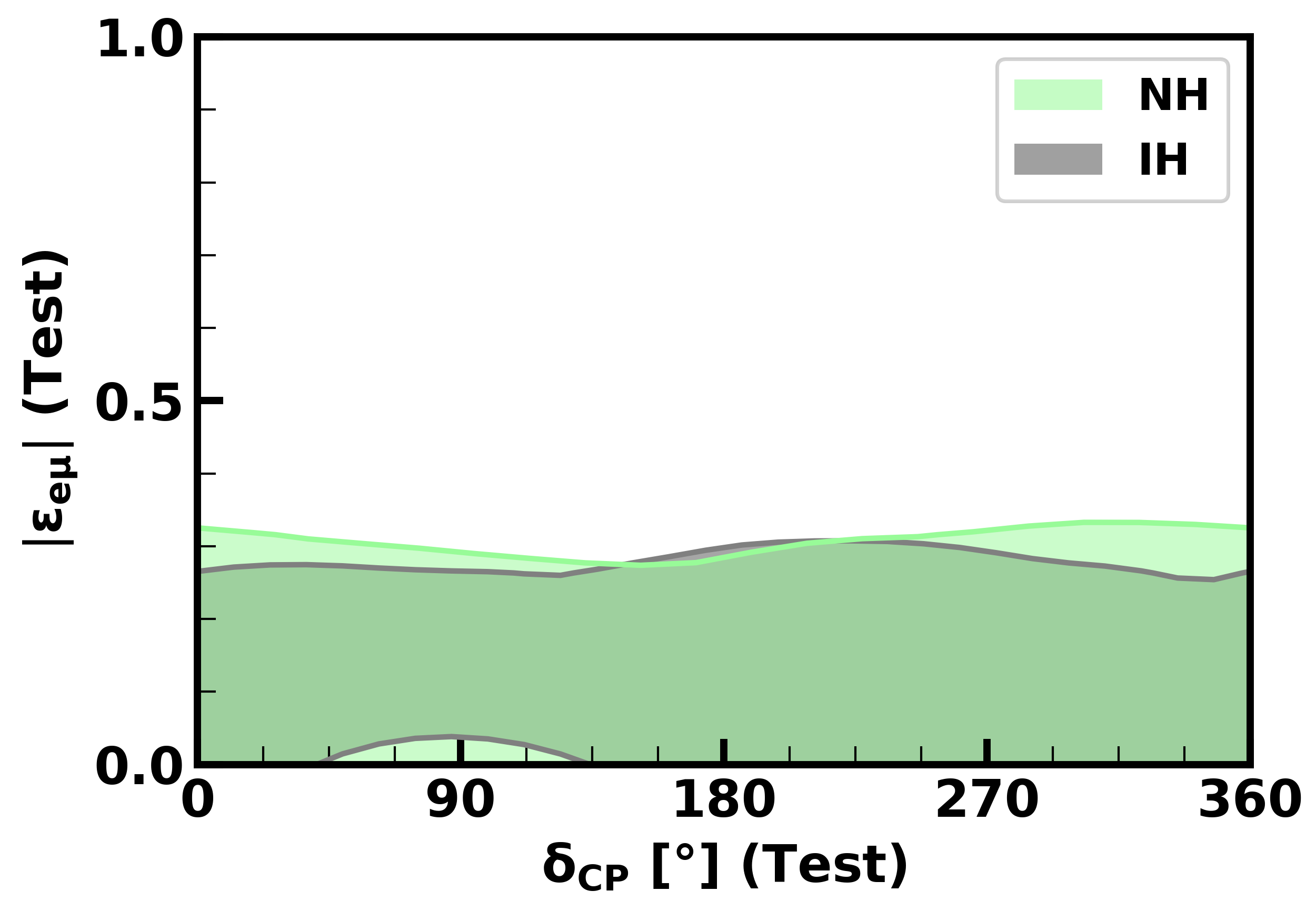}
\includegraphics[width=0.45\linewidth]{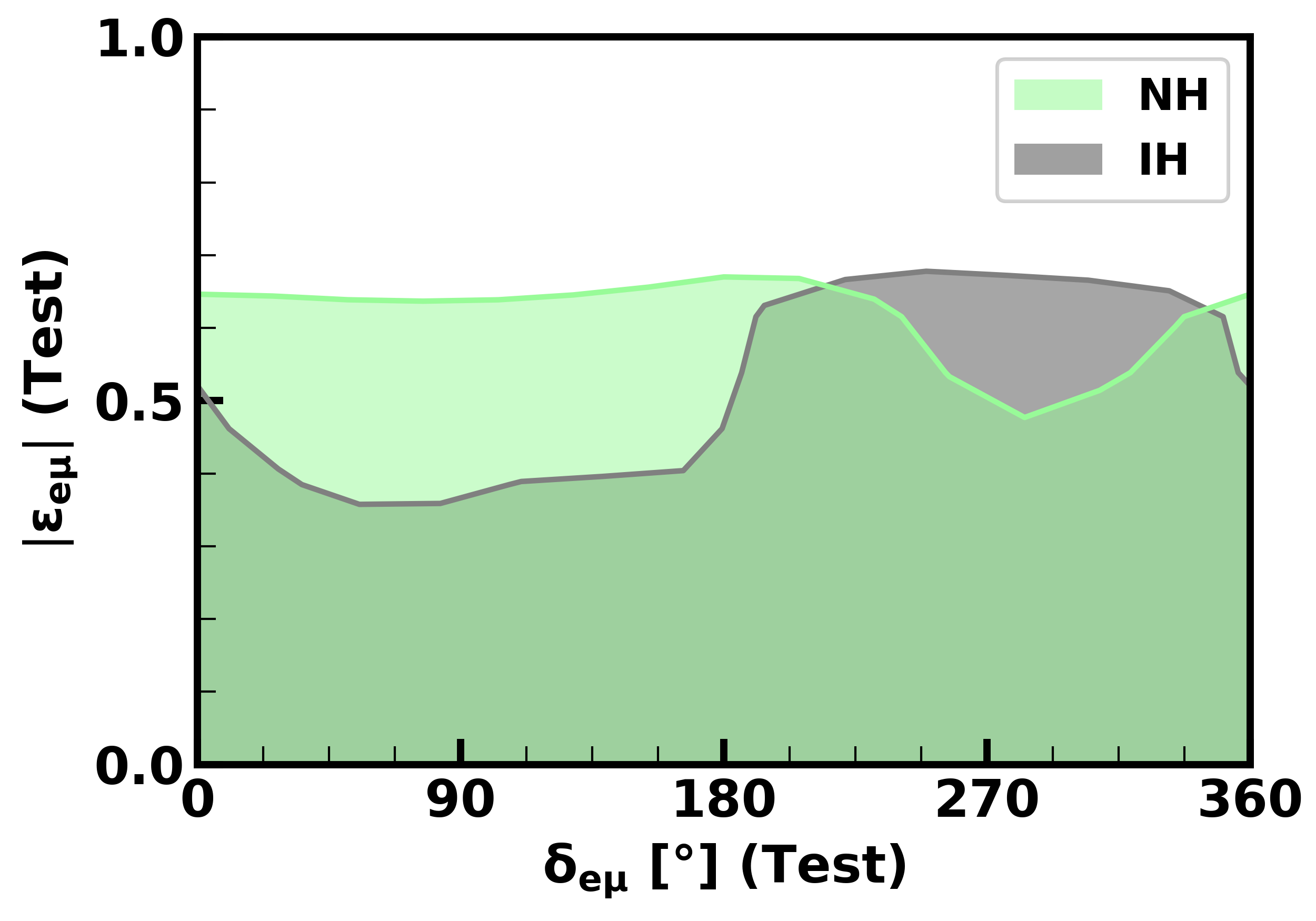}
\includegraphics[width=0.45\linewidth]{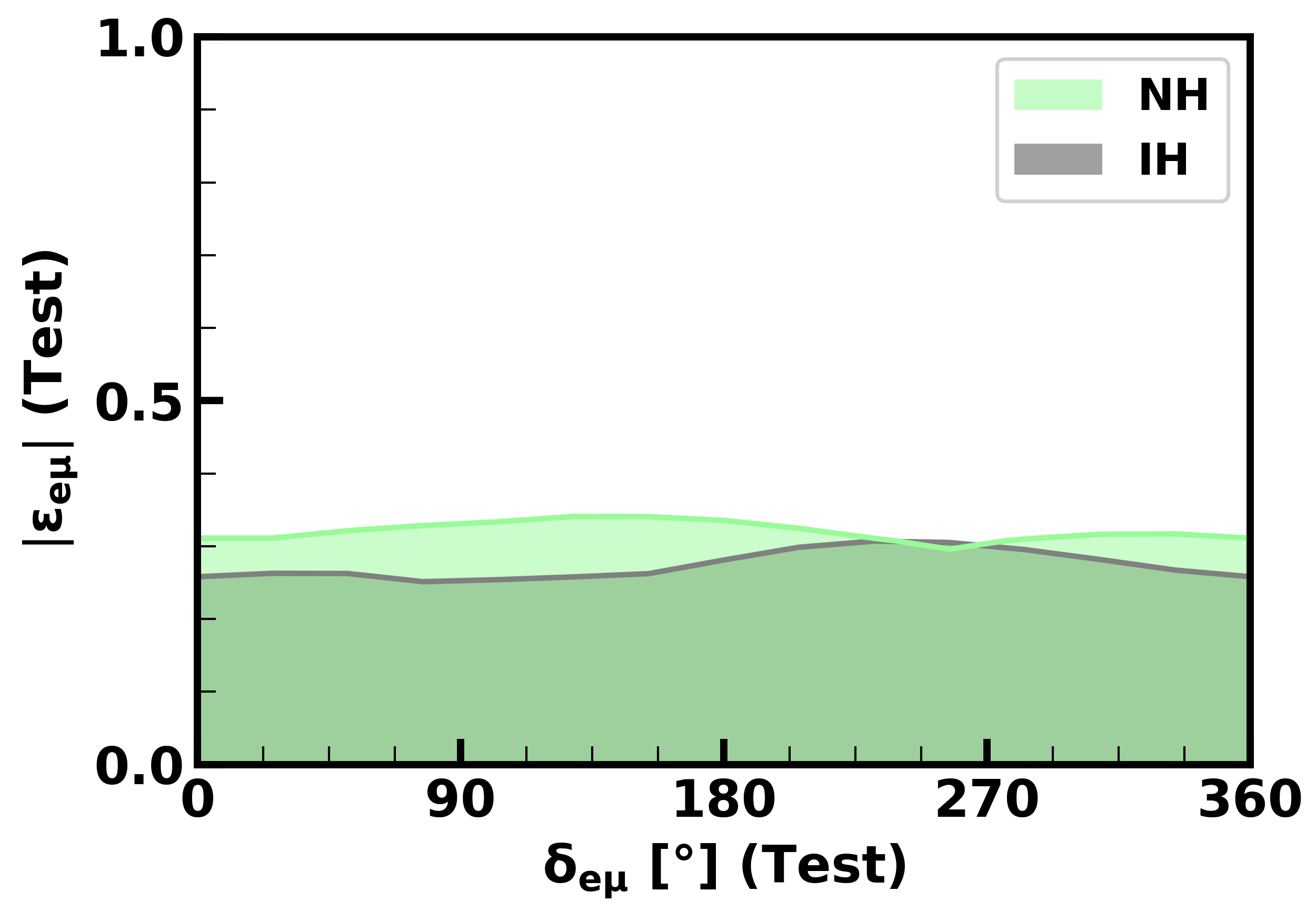}
\caption{$\epsilon_{e\mu}$ (test) vs $\delta_{CP}$ (test) in the upper row and $\epsilon_{e\mu}$ (test) vs $\delta_{e\mu}$ in the lower row. In the left panel 1 - 5 GeV and in the right panel 1 - 20 GeV. Marginalized over $\theta_{13}$, $\theta_{23}$, $\Delta m^2_{31}$, $\delta_{e\mu}$ ($\delta_{CP}$) in upper row (lower row).}
\label{fig:eps_emu-vs-delcp_del_emu}
\end{figure*}

In fig.~\ref{fig:eps_emu-vs-delcp_del_emu} we show 2-dimensional sensitivity analysis by projecting $\chi^2$ in $|\epsilon_{e\mu}|(test) - \delta_{CP}(test)$ plane (upper row) and $|\epsilon_{e\mu}|(test) - \delta_{e\mu}(test)$ plane (lower row). We plot the contours in the left and right panels considering simulated NOvA events for standard framework ($1<E_\nu<4$ GeV) and including high energy ($1<E_\nu<20$ GeV), respectively. We calculate $\Delta \chi^2$ hypothesizing standard oscillation in the true simulated data and NSI in the theory and plot $95\%$ CL contour in the $\epsilon_{e\mu}$[test] versus $\delta_{CP}$[test] plane. We define the $\Delta \chi^2$ as
\begin{equation}
    \Delta \chi^2 = \chi^2 \left(\epsilon_{e\mu} [true] = 0, \epsilon_{e\mu} [test] \neq 0 \right) ,
    \label{eq:chi-sq-eps_emu}
\end{equation}
for the fix true $\delta_{CP}$ at $0.82\pi$ and marginalize over $\theta_{13},~\theta_{23},~\Delta m^2_{31}$ and $\delta_{e\mu}$. 
Similarly, we calculate $\Delta \chi^2$ using eq.~\ref{eq:chi-sq-eps_emu} corresponding to  $\epsilon_{e\mu}$[test] versus $\delta_{e\mu}$[test] contour, where we marginalize over $\theta_{13},~\theta_{23},~\Delta m^2_{31}$ and $\delta_{CP}$. We vary $\delta_{CP}$ and $\delta_{e\mu}$ in their full range $0-2\pi$ and show contours with $95\%$ CL for both the normal hierarchy (NH, green regions) and inverted hierarchy (IH, gray region). The unmentioned oscillation parameters are kept fixed to their true values. Note that NOvA disfavors the region above the gray line for NH and green line for IH.

From the top-left plot, considering NH, we observe the allowed region (with $95\%$ CL) favors relatively lower $|\epsilon_{e\mu}|$ in the $\delta_{CP} \sim [90\degree:220\degree]$, whereas the contour favors higher $\epsilon_{e\mu}$ elsewhere. For IH, the contour encompasses relatively higher $|\epsilon_{e\mu}|$ in the $\delta_{CP} \sim [90\degree:270\degree]$, while it adopts lower $\epsilon_{e\mu}$ elsewhere. In the bottom-left plot, considering NH (IH), the $95\%$ CL contour favors relatively lower (higher) $|\epsilon_{e\mu}|$ in the $\delta_{e\mu} \sim [220\degree : 340\degree]~([180\degree : 360\degree])$. The range of $\epsilon_{e\mu}$ corresponding to allowed contours are identical in both the upper and lower plots as well as for NH and IH. This implies that the sensitivity to $\epsilon_{e\mu}$ remains almost unchanged w.r.t $\delta_{CP}$ or $\delta_{e\mu}$ and irrespective of mass hierarchy.
A noteworthy fact is that the contours favor relatively lower $\epsilon_{e\mu}$ in the presence of high energy events for both the plots in the right panel. This indicates, including high energy events provides better constraints on $\epsilon_{e\mu}$.


\section{Study of environmental decoherence including $\nu_e / \overline{\nu}_e$ high energy events}\label{sec:env-deco}

Neutrino oscillation data from the experiments rely on the three flavor mixing where the propagation of neutrinos is interpreted as the coherent superposition of mass eigenstates. In this quantum evolution the degree of coherence remains unaffected over the macroscopic distances because of the weak coupling of neutrinos with matter. Such phenomenon manifests the isolation of neutrino systems from their surroundings while propagation.
Nevertheless, neutrinos could interact with the stochastic environment~\cite{Stuttard:2020qfv}, for instance fluctuating space-time at the Planck scale, matter density fluctuation, virtual black hole. This incorporates the loss of coherence in propagating neutrinos states and introduces a non-standard effect which is known as environmental decoherence\footnote{This effect is different from neutrino wave-packet decoherence~\cite{Giunti:1998kim, Blennow:2005ohl, Akhmedov:2012her, Chang:2016chu, Gouvea:2021rom}}. In this phenomenology we assume neutrino (sub-)system is an open quantum system and interacts with the environment. In the presence of environmental interaction the neutrino oscillation probability gets modified and carries damping term $e^{-\Gamma L}$, where $\Gamma$ is the decoherence parameter. $\Gamma$ could depend on the different power of the neutrino energy as $\Gamma \propto E^n$, here choice of $n$ is based on the various physical origins of decoherence phenomenon.

\subsection{Mathematical formulation of oscillation probabilities}
\label{theory:deco}
Neutrino propagation in  an open quantum system framework introduces loss of coherence due to interaction of neutrino subsystem with the stochastic environment. This loss is incorporated using Lindblad master equation which represents the density matrix evolution as given below~\cite{Lindblad:1976g, GKS:1976vit}
\begin{equation}
    \frac{d\rho^m(t)}{dt} = -i\left [H,\rho^m(t) \right] + \mathcal{D} \left[ \rho^m (t) \right],
    \label{eq:LME}
\end{equation}
where $\rho_m$ is the density matrix of neutrino states in the mass basis and $H$ is the Hamiltonian of the neutrino system. The damping term $\mathcal{D}[\rho_m (t)]$ represents the interaction neutrino system and the environment. The parameterization of $\mathcal{D}[\rho_m (t)]$ is obtained by using complete positivity and trace preserving conditions. Applying complete positivity modifies the Lindblad form of the dissipator as~\cite{GKS:1978g} 
\begin{equation}
\begin{aligned}
    \mathcal{D} \left[ \rho^m (t) \right] &= \frac{1}{2}\sum_{n = 1}^{N^2 - 1} \left\{[\mathcal{V}_n , \rho^m \mathcal{V}_n^\dagger] + [\mathcal{V}_n \rho^m , \mathcal{V}_n^\dagger]\right\},
    \end{aligned}
        \label{eq:D-term}
\end{equation}
\noindent
here $N$ is the dimension of the Hilbert space and $\mathcal{V}_n$ are the interaction operators. Further, we impose an increase in von Neumann entropy $ S = - Tr(\rho^m \ln \rho^m) $~\cite{Banks:1984bsp,Benatti:1988nar} and conservation of average energy $Tr(\rho^m H)$ of the neutrino system and obtain the elements of time evolved density matrix,
\begin{equation}
\begin{aligned}
    & \rho^m_{jk}(t) = \rho_{jk}(0),~for~j = k,\\
    & \rho^m_{jk}(t) = \rho_{jk}(0) \exp-(\Gamma_{jk} + i \Delta_{jk}),~for~j \neq k.
    \end{aligned}
\end{equation}
with 
\begin{equation}
    \Gamma_{jk} = \Gamma_{jk} = \frac{1}{2}\sum_{n = 1}^8 (d_{n,j} - d_{n,k})^2
    \label{eq:Gamma_ij},
\end{equation}
where, $d_{n,j}$, $d_{n,k}$ are the diagonal elements of $\mathcal{V}_n$ operator and $j,k$ take the values $1,2,3$.

We consider the modified-mixing matrix ($\tilde{U}$) up to the 1st order approximation, from ref.~\cite{Denton:2018dmp} to convert mass basis to flavor basis using $\tilde{\rho}^{\alpha} = \tilde{U}~\tilde{\rho}^m~\tilde{U}^\dagger$. The neutrino transition probability from initial flavor '$\nu_\alpha$' to final flavor '$\nu_\beta$' in terms of density matrix is obtained by
\begin{equation}
   \begin{aligned}
        P_{\alpha \beta}(t) &= Tr[\tilde{\rho}_\alpha (t) \tilde{\rho}_\beta (0)].
   \end{aligned}
   \label{eq:P1}
\end{equation}
The explicit form of the transition probability assuming ultra-relativistic neutrinos ($t \approx L$) is given by~\cite{Gomes:2019for,Coloma:2018ice}
\begin{equation}
\begin{aligned}
        P_{\alpha \beta}(L) &= \delta_{\alpha \beta} - 2\sum_{j > k} Re \left( \tilde{U}_{\beta j} \tilde{U}_{\alpha j}^* \tilde{U}_{\alpha k} \tilde{U}_{\beta k}^* \right) + 2\sum_{j > k} Re \left( \tilde{U}_{\beta j} \tilde{U}_{\alpha j}^* \tilde{U}_{\alpha k} \tilde{U}_{\beta k}^* \right) \exp(-\Gamma_{jk} L) \cos(\frac{\tilde{\Delta}m_{jk}^2}{2E}L) \\& + 2\sum_{j > k} Im \left( \tilde{U}_{\beta j} \tilde{U}_{\alpha j}^* \tilde{U}_{\alpha k} \tilde{U}_{\beta k}^* \right) \exp(-\Gamma_{jk} L) \sin(\frac{\tilde{\Delta}m_{jk}^2}{2E}L)~.
        \end{aligned}
        \label{eq:Pab}
\end{equation}

Here in eq.~(\ref{eq:Pab}), the term $e^{-\Gamma_{jk}L}$ damps the oscillation probability. We analyze the effect of decoherence by considering all the $\Gamma_{jk}$ as equal (i.e., $\Gamma_{21} = \Gamma_{31} = \Gamma_{32}$) and represent it by a single parameter $\Gamma$ for simplicity.
Further we assume a general power law dependency of $\Gamma$ parameter on the neutrino energy given by 
\begin{equation}
    \Gamma(E_\nu) = \Gamma_0\left(\frac{E_\nu}{E_0}\right)^n~,
    \label{G-powerlaw}
\end{equation}
where, $\Gamma_0$ is constant, $E_0$ is the reference energy taken as 1 GeV and $n = 0,\pm 1,\pm 2$. 
Different physical origins explain the decoherence phenomena leading to different integral power law dependencies~\cite{Romeri:2023cgt}. However, from eq.~(\ref{G-powerlaw}) one can note that the oscillation probabilities are sensitive to $n>0$ when $E_\nu > E_0$. In case of NOvA experiment $E_\nu \sim 1.8$ GeV is greater than the reference energy (1 GeV), so environmental decoherence plays significant role when $n > 0$ and which could yield stronger constraints on $\Gamma$.

It is noteworthy to mention that, the neutrino oscillation probabilities are sensitive to positive powers of $n$ (leftmost panel of fig.~\ref{fig:decoherence-prob-event}). This happens because the energy of the neutrino beam $E_\nu$ is greater than the reference energy $E_0 = 1$ GeV ($E_\nu \geq E_0$) in eq.~\ref{G-powerlaw}.

The analytical appearance and disappearance probabilities considering perturbative expansion up to the first order in small parameters $s_{13}$ and $\eta$ are shown below
\begin{equation}\label{eq:app-prob-approx-deco}
    \begin{aligned}
        P_{\mu e} =& \frac{2}{\hat{A} - 1}\left[\frac{\eta s_{12} c_{12} s_{13} s_{23} c_{23}}{\hat{A}}\left\{-e^{-\Gamma_{21}L}\cos{(\frac{\Delta m^2_{21}L}{2E})} - e^{-\Gamma_{31}L}\cos{(\frac{\Delta m^2_{31}L}{2E})} \right. \right.\\
        &\left. \left. + e^{-\Gamma_{32}L}\cos{(\frac{\Delta m^2_{32}L}{2E})} + 1 \right\}  - \frac{\hat{A} - 1}{\hat{A}^2} \eta^2 s_{12}^2 c_{12}^2 c_{23}^2 \left\{ e^{-\Gamma_{21}L}\cos{(\frac{\Delta m^2_{21}L}{2E})} - 1 \right\} \right.\\
        &\left. -\frac{s_{13}^2 s_{23}^2}{\hat{A} - 1} \left\{ e^{-\Gamma_{31}L}\cos{(\frac{\Delta m^2_{31}L}{2E})} - 1 \right\} \right]~,
    \end{aligned}
\end{equation}
and 
\begin{equation}\label{eq:disapp-prob-approx-deco}
    \begin{aligned}
        P_{\mu\mu} =& 1 + 2 \left[ \frac{c_{23}^2}{\hat{A}^2} \eta^2 s_{12}^2 c_{12}^2 + \frac{2 c_{23} s_{23}}{\hat{A}(\hat{A}-1)}  \eta s_{12} c_{12} s_{13} + c_{23}^2 s_{23}^2 \left\{ e^{-\Gamma_{32}L}\cos{(\frac{\Delta m^2_{32}L}{2E})} -1 \right\} +  \right.\\ &\left. + \frac{s_{23}^2}{(\hat{A}-1)^2}  s_{13}^2 
       \left\{ c_{23}^2 e^{-\Gamma_{21}L}\cos{(\frac{\Delta m^2_{21}L}{2E})} + s_{23}^2 e^{-\Gamma_{31}L}\cos{(\frac{\Delta m^2_{31}L}{2E})} -1 \right\}  \right]~.
    \end{aligned}
\end{equation}


\subsection{Oscillation probabilities and event rates}

\begin{figure*}[!htbp]
\includegraphics[width=0.32\linewidth]{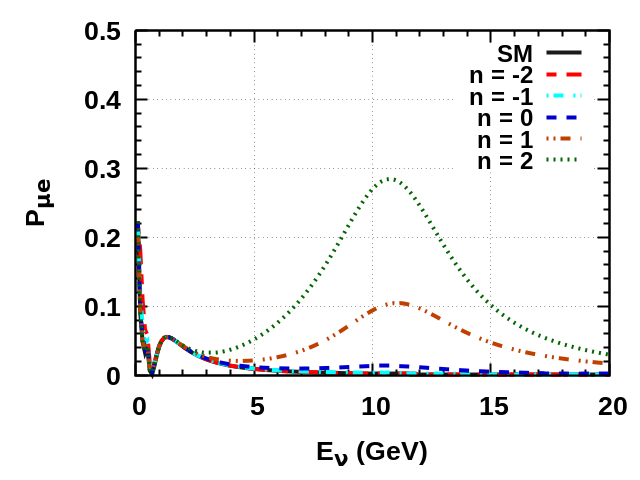}
\includegraphics[width=0.32\linewidth]{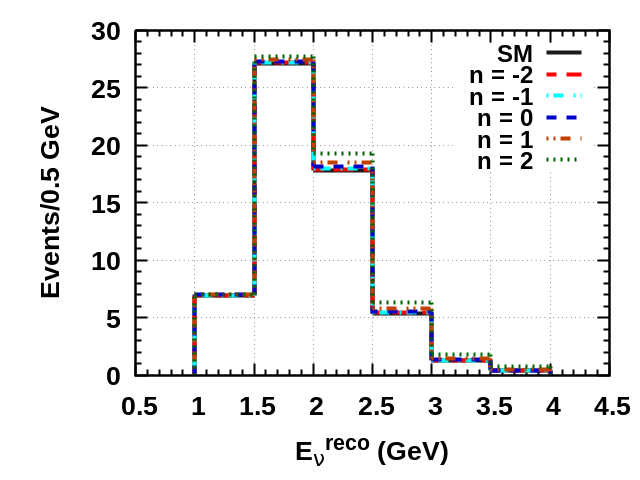}
\includegraphics[width=0.32\linewidth]{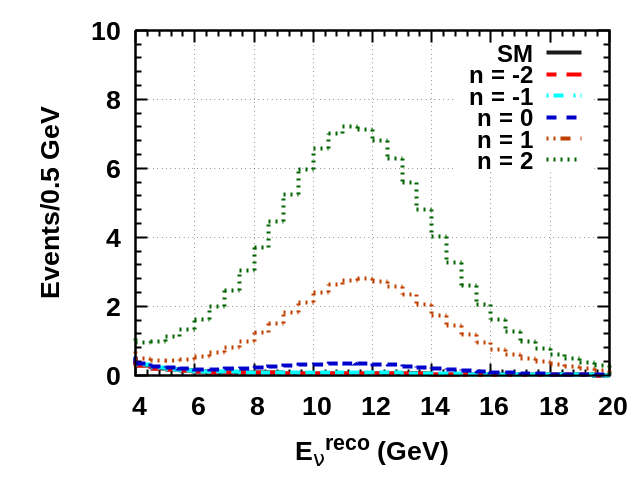}
\includegraphics[width=0.32\linewidth]{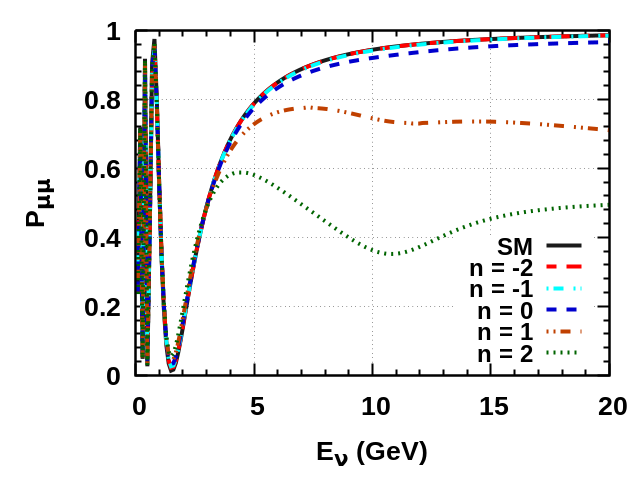}
\includegraphics[width=0.32\linewidth]{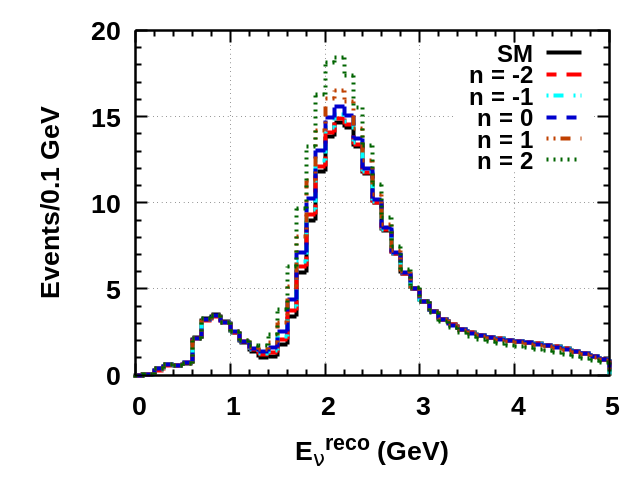}
\includegraphics[width=0.32\linewidth]{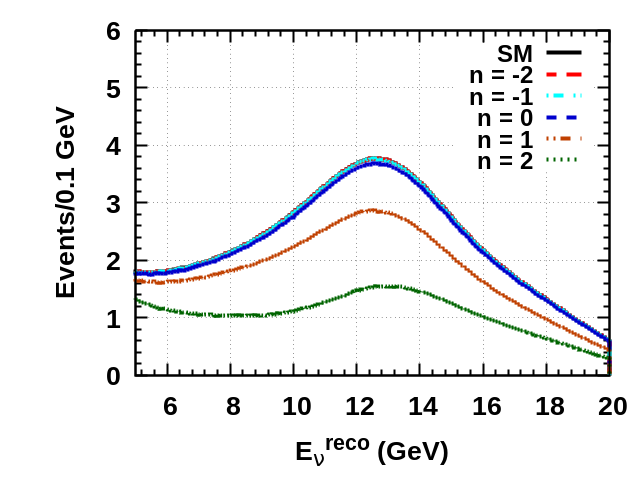}
\caption{$\nu_e$ appearance probability and event rates in the top row. $\nu_\mu$ disappearance probability and event events in the bottom row. We assume $\Gamma = 10^{-23}$ GeV.}
\label{fig:decoherence-prob-event}
\end{figure*}

We show the oscillation probability in the presence of decoherence effect (using eq.~(\ref{eq:Pab})) to demonstrate the deviation as compared to standard three flavor scenarios. In the top row of fig.~\ref{fig:decoherence-prob-event} we depict $\nu_e$-appearance probability and corresponding event rates, and in the bottom row $\nu_\mu$-disappearance probability and corresponding event rates. In the left panel we show probabilities as a function of energy. In the middle and right panels we illustrate event rates as a function of reconstructed energy correspond to $1<E_\nu<4$ GeV and $4 < E_\nu < 20$ GeV, respectively. We compute the probabilities and event rates numerically considering standard parameters from table~\ref{table:1} and assuming decoherence parameter $\Gamma = 10^{-23}$ GeV along with the energy power-law dependencies. In each plot we display the curves corresponding to the standard oscillation and decoherence with power-law indices $n = 0,\pm 1,\pm 2$ as per their respective legends shown in the plots. We tabulate the event rates for all the mentioned cases in table~\ref{table:deco_sig_events}.


From the left plots we observe a marginal impact of decoherence on the oscillation probabilities in the energy range $1 < E_\nu < 4$ GeV.  Adopting power-law indices $n = -2,-1$ (red, cyan curves) we notice negligible effect of $\Gamma$ on the appearance and disappearance channels in the energy window $4<E_\nu<20$ GeV. For $n = 0$ (blue) we see the effect of decoherence is less prominent, whereas $n = 1,2$ lead to a noticeable impact on both the appearance and disappearance probabilities. Additionally $\sim 11$ GeV we see an increase in $\nu_\mu \rightarrow \nu_e$ probability (top left) and decrease in $\nu_\mu \rightarrow \nu_\mu$ (bottom left) as compared to the corresponding standard oscillation scenario. 
From these plots we observe that the neutrino oscillation probabilities are deviate from SM case for $n>0$. This is the consequence of power-law in eq.~(\ref{G-powerlaw}), since our analysis includes neutrino energy higher than the reference energy i.e., ($E_0 = 1$ GeV). The details of peak that arises around $\sim 11$ GeV has been discussed in ref.~\cite{Bera:2024hhr}. The event rates in middle and right panels also follow a similar trend as the oscillation probabilities. For power-law indices $n = 0,1,2$~\footnote{Energy independent decoherence when $n = 0$, whereas $n = 1$ refers to the linear energy dependence that could be modeled from \textit{foamy} fluctuations in space-time~\cite{Ellis:1999uh}, and $n = 2$ leads to the quadratic energy dependency on $\Gamma$ could be induced by quantum gravity effect~\cite{Ellis:1997jw}.} considering HE events could provide better constraints on $\Gamma$.

\begin{table}
\setlength{\tabcolsep}{5pt}
  \begin{tabular}{|l|l|l|l|l|l|l|l|}
    \hline
    \multicolumn{8}{|c|}{Signal events} \\
    \hline
    \multicolumn{2}{|c|}{\multirow{1}{*}{Channel}} &
      \multicolumn{3}{c|}{$\nu_\mu \rightarrow \nu_e$} &
      \multicolumn{3}{c|}{$\nu_\mu \rightarrow \nu_\mu$} \\
      \cline{1-8}
      \multicolumn{2}{|c|}{\multirow{1}{*}{Energy range (GeV)}} & $1 - 4$ & $1 - 20$ & $4 - 20$ & $1 - 4$ & $1 - 20$ & $4 - 20$\\
      \hline
     \multicolumn{2}{|c|}{SM} & 59.0 & 61.0 & 2.0 & 215.0 & 589.8 & 374.8\\
     \hline
     {\multirow{5}{*}{$\Gamma_0 = 10^{-23}$ GeV}} & $n = -2$ & 59.1 & 61.2 & 2.1 & 218.0 & 592.7 & 374.7\\
     \cline{2-8}
     & $n = -1$ & 59.3 & 61.7 & 2.4 & 220.8 & 594.9 & 374.1\\
     \cline{2-8}
    & $n = 0$ & 59.7 & 66.0 & 6.3 & 225.9 & 592.3 & 366.4\\
     \cline{2-8}
    & $n = 1$ & 60.7 & 101.0 & 40.3 & 234.8 & 529.8 & 295.0\\
     \cline{2-8}
    & $n = 2$ & 63.0 & 165.9 & 102.9 & 250.1 & 423.8 & 173.7\\
    
    \hline
  \end{tabular}
  \caption{Excess number of events in HE ($4<E_\nu<20$ GeV) considering $\Gamma \neq 0$~.}
  \label{table:deco_sig_events}
\end{table}

\subsection{Upper bounds on the decoherence parameter $\Gamma$}

To obtain the upper bounds on $\Gamma_0$ we show 2-dimensional $\chi^2$ distribution in the test $\Gamma_0$ versus true $\delta_{CP}$ plane considering power-law indices $n = 0,1,2$. Here we are not showing the corresponding plots for $n = -2, -1$ as the effect of $\Gamma$ on $P_{\mu e}$ and $P_{\mu \mu}$ is negligible as seen in fig.~\ref{fig:decoherence-prob-event}.  We compute the $\chi^2$ values for every true $\delta_{CP} \in [0,360\degree]$ by assuming standard oscillation phenomenon in the simulated data and the decoherence scenario in the theory. We provide the definition of $\chi^2$ below
\begin{equation}\label{eq:chi-Gamma}
    \chi_{\Gamma}^2 (\delta_{CP}^{true}) = \chi^2 (\Gamma^{true} = 0, \Gamma^{test} \neq 0)~,
\end{equation} 
and marginalize over test parameters $\delta_{CP}$, $\theta_{23}$, $\Delta m^2_{31}$. 

\begin{figure*}[!htbp]
\includegraphics[width=0.32\linewidth]{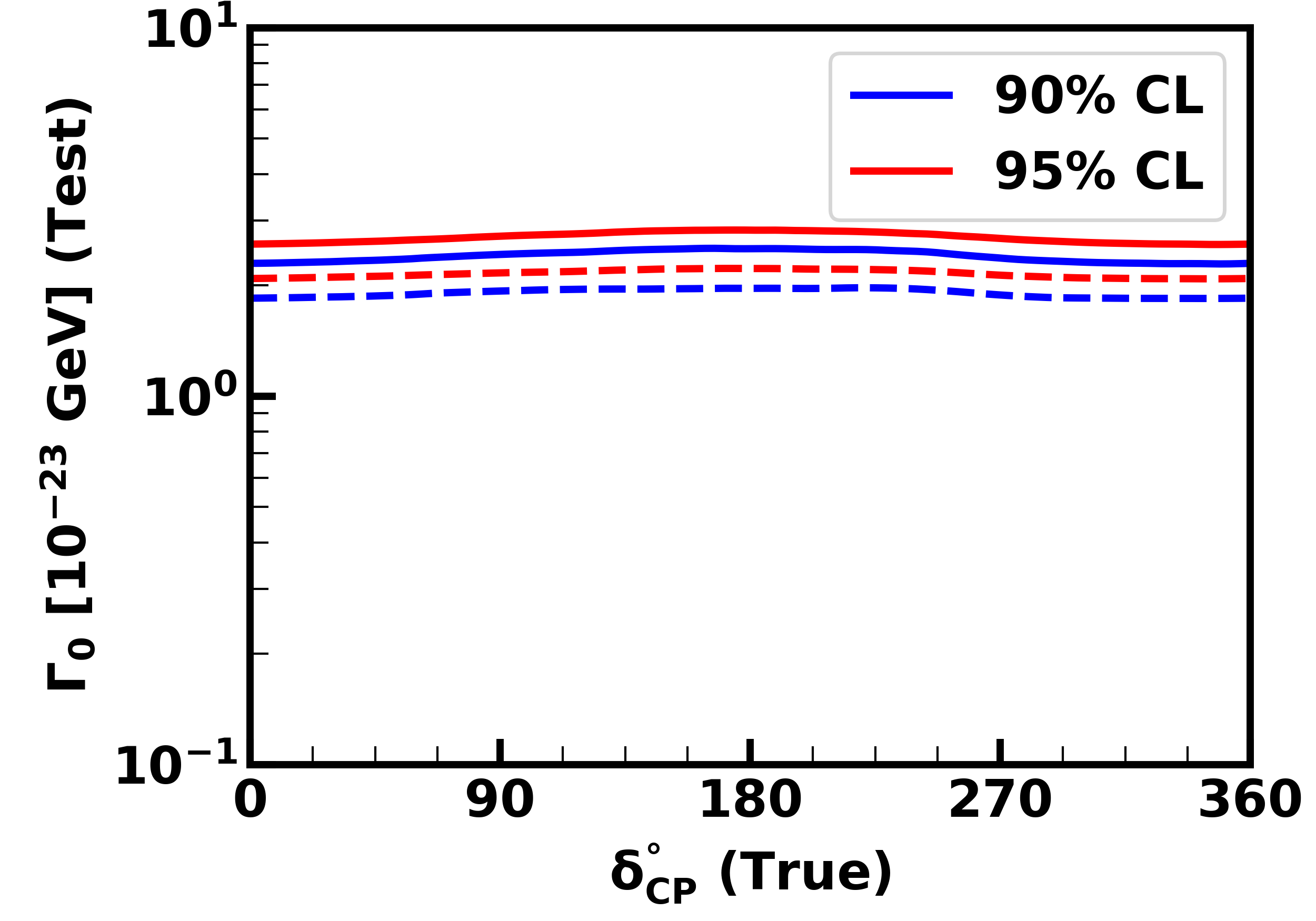}
\includegraphics[width=0.32\linewidth]{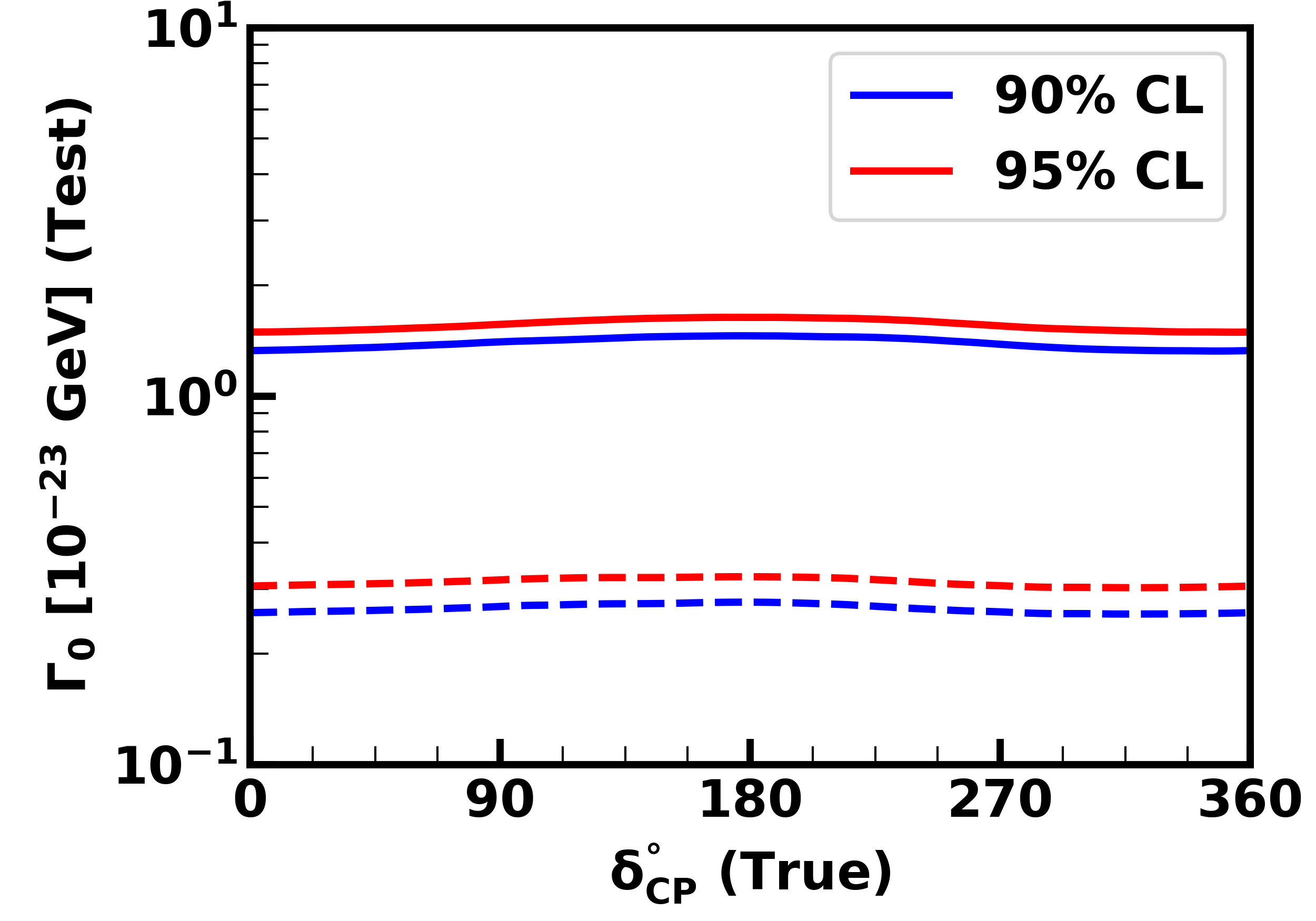}
\includegraphics[width=0.32\linewidth]{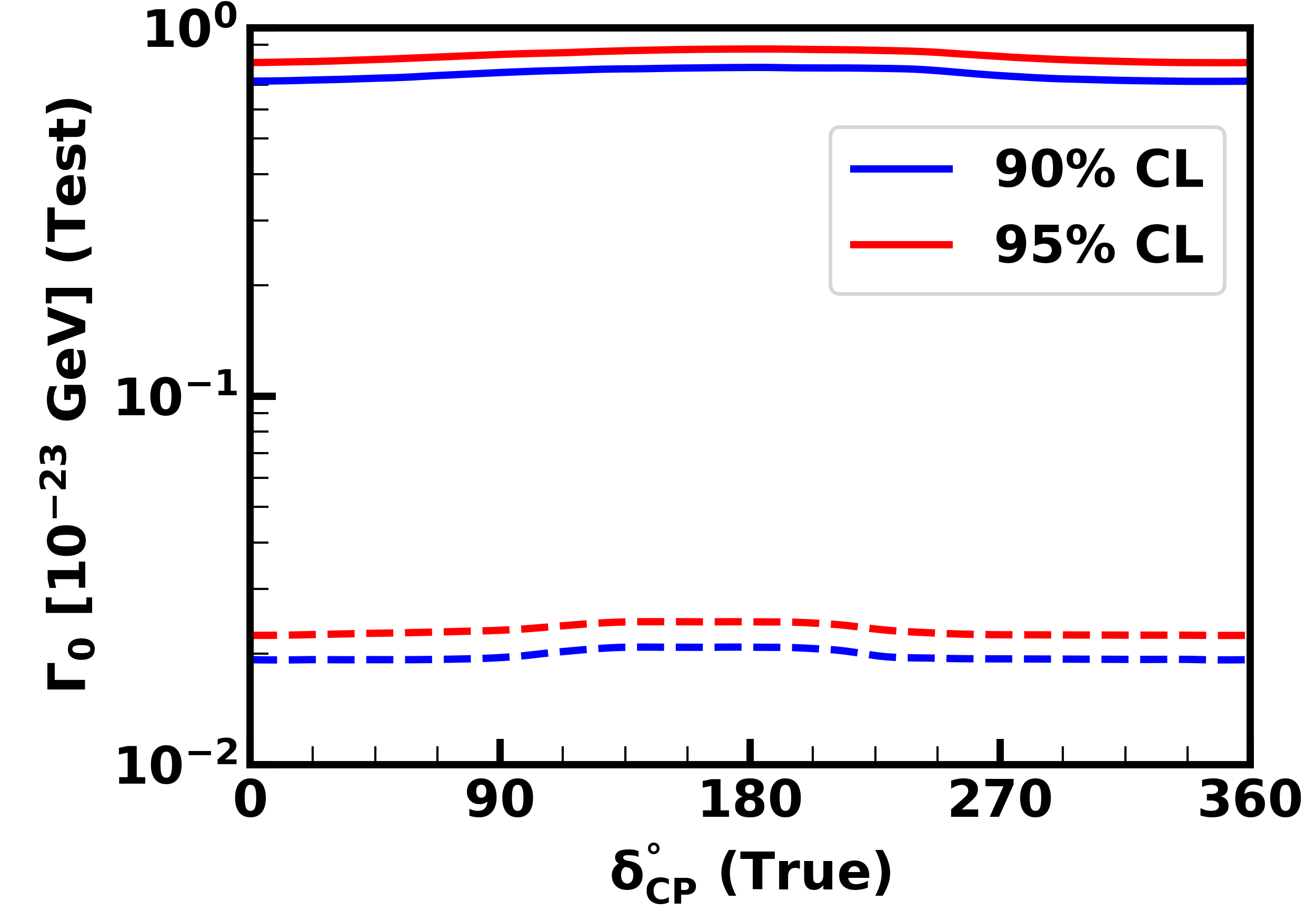}
\caption{Constraining $\Gamma_0$ using energy range $1 - 4$ GeV (solid lines) and $1- 20$ GeV (dashed lines). Left, middle and right plots correspond to $n = 0$, $n = 1$ and $n = 2$ respectively. We marginalize over $\Delta m^2_{31}$, $\theta_{23}$, $\delta_{CP}$.}
\label{fig:bounds-deco-params}
\end{figure*}

In fig.~\ref{fig:bounds-deco-params} the left, right and middle plots represent the results for $n=0$, $n=1$ and $n=2$ respectively. In each of these plots the solid lines depict the $\chi^2$ obtained for the energy range $E_\nu : 1-4$~GeV while the dashed lines show the results for $E_\nu : 1-20$~GeV. 
The blue and red curves stand for $90\%$ and $95\%$ CL respectively. 

Across all the plots we point out from the solid and dashed curves that including high energy events ($E_\nu : 1-20$~GeV) provide comparatively better constraint on $\Gamma_{jk}$ than NOvA standard analysis window $E_\nu : 1-4$~GeV. However, this confirmation is less significant assuming $n=0$ (left) and dominate with increasing $n$ ($=1,2$; middle, right plots). We can explain such behavior from the $\nu_e$-appearance probability and corresponding event rates in fig.~\ref{fig:decoherence-prob-event}. In all the plots we see that $90\%$ (blue) and $95\%$ CL (red) curves are closer to each other therefore we list the bounds corresponding to $90\%$ CL in this section. We also notice that the corresponding bounds on $\Gamma$ remain almost constant (as blue curves are flat) for all values of $\delta_{CP}$. This indicates that the decoherence parameter $\Gamma$ does not depend on the true value of $\delta_{CP}$.

 Considering $E_\nu : 1-4$~GeV the sensitivity at $90\%$ CL to the $\Gamma_0$ for $n = 0,1,2$ is listed as $2.2 \times 10^{-23}$ GeV, $1.3 \times 10^{-23}$ GeV, $7.15 \times 10^{-24}$ GeV respectively. Whereas analyzing events from $E_\nu : 1-20$~GeV the sensitivity at $90\%$ CL to the $\Gamma_0$ for $n = 0,1,2$ is listed as $1.82 \times 10^{-23}$ GeV, $2.52 \times 10^{-24}$ GeV, $2.0 \times 10^{-25}$ GeV respectively. Clearly, for $n= 1, 2$ we note more stringent bounds (by $\mathcal{O}(1)$) when we added high energy events from $E_\nu : 4-20$~GeV. Notably, in the case of $n = 2$ (right most plot) we see a significant improvement in the bounds of $\Gamma$ in comparison to the results obtained from assuming $E_\nu : 1-4$~GeV and this can be observed from solid and dashed curves. To facilitate a better comparison we list all the bounds in the table~\ref{table:bounds-on-gamma}. 

 \begin{table}
\setlength{\tabcolsep}{5pt}
  \begin{tabular}{|l|lll|}
  \hline
    \multicolumn{1}{|c|}{\multirow{2}{*}{Analysis window}} & \multicolumn{3}{c|}{Bounds on $\Gamma_0$ (GeV)} \\
    \multicolumn{1}{|c|}{} & $n = 0$ & $n = 1$ & $n = 2$ \\
    \hline
    $E_\nu : 1-4$ GeV & $2.2 \times 10^{-23}$ & $1.3 \times 10^{-23}$ & $7.15 \times 10^{-24}$ \\
    $E_\nu : 1-20$ GeV & $1.82 \times 10^{-23}$ & $2.52 \times 10^{-24}$ & $2.0 \times 10^{-25}$ \\
    \hline
  \end{tabular}
  \caption{Bounds on $\Gamma_0$ obtained from fig~\ref{fig:bounds-deco-params} for different power-law indices $n = 0,1,2$.}
  \label{table:bounds-on-gamma}
\end{table}


\subsection{Allowed regions in $\theta_{23}$-$\delta_{CP}$ plane}

\begin{figure*}[!htbp]
\includegraphics[width=0.45\linewidth]{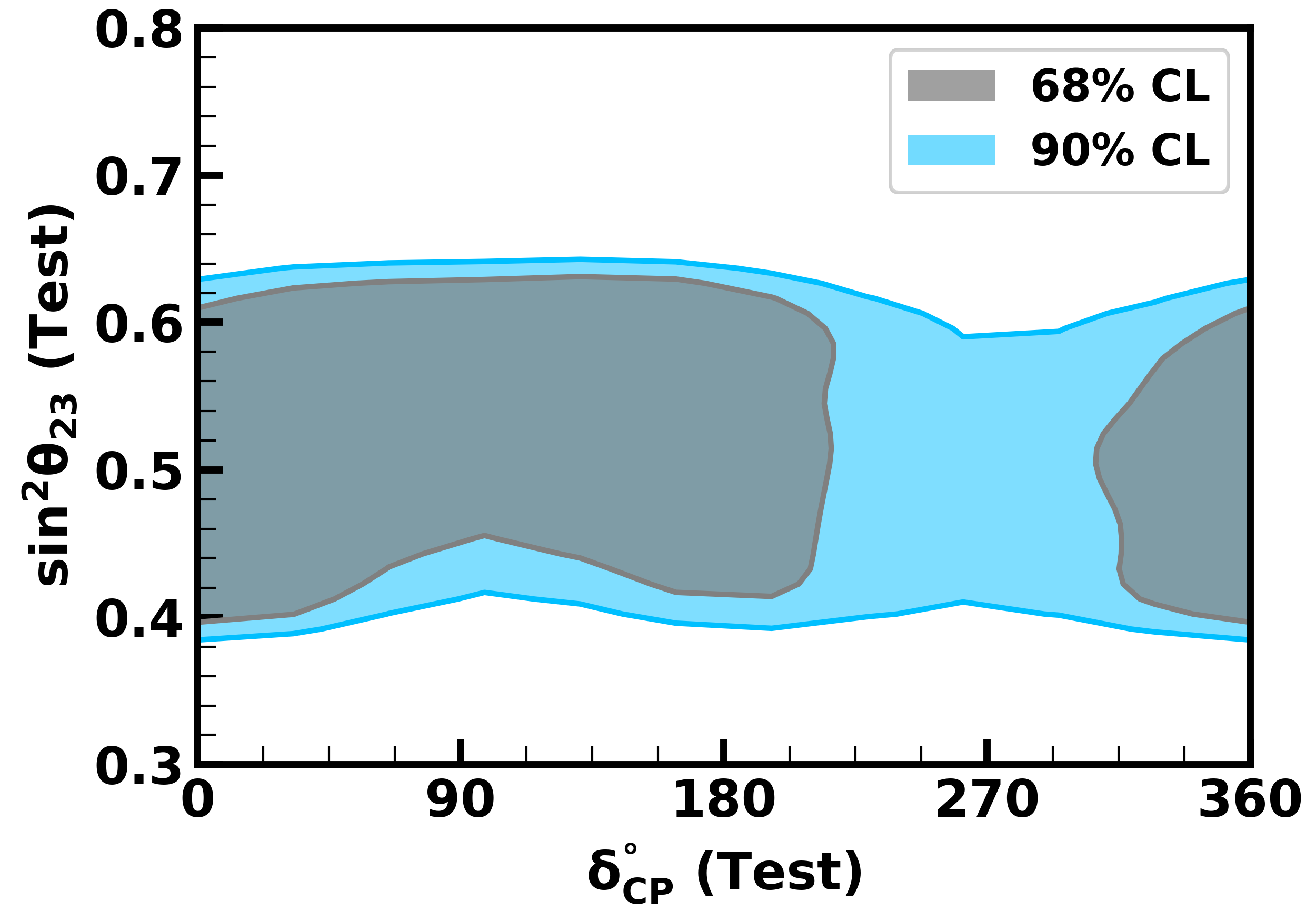}
\includegraphics[width=0.45\linewidth]{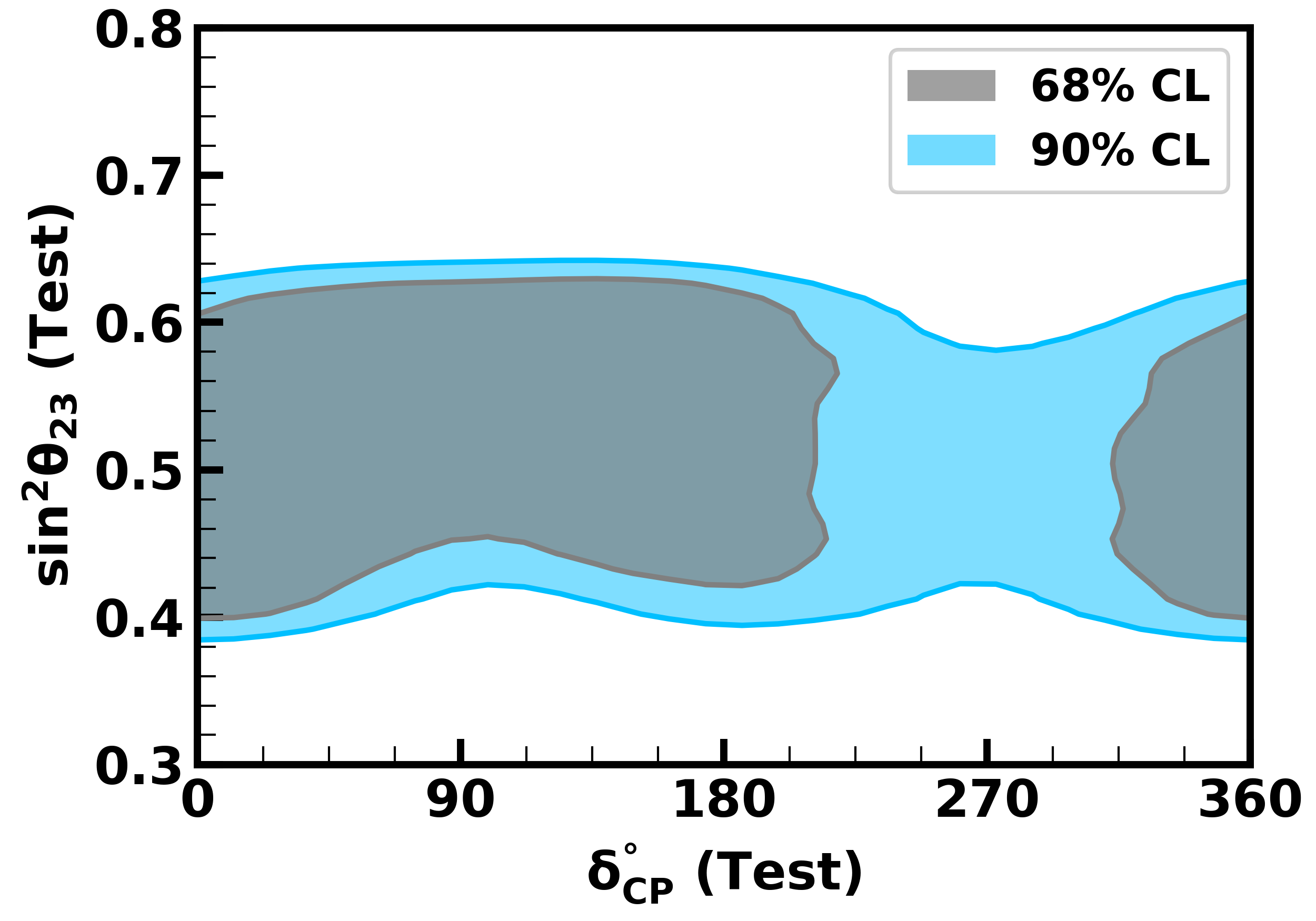}
\includegraphics[width=0.45\linewidth]{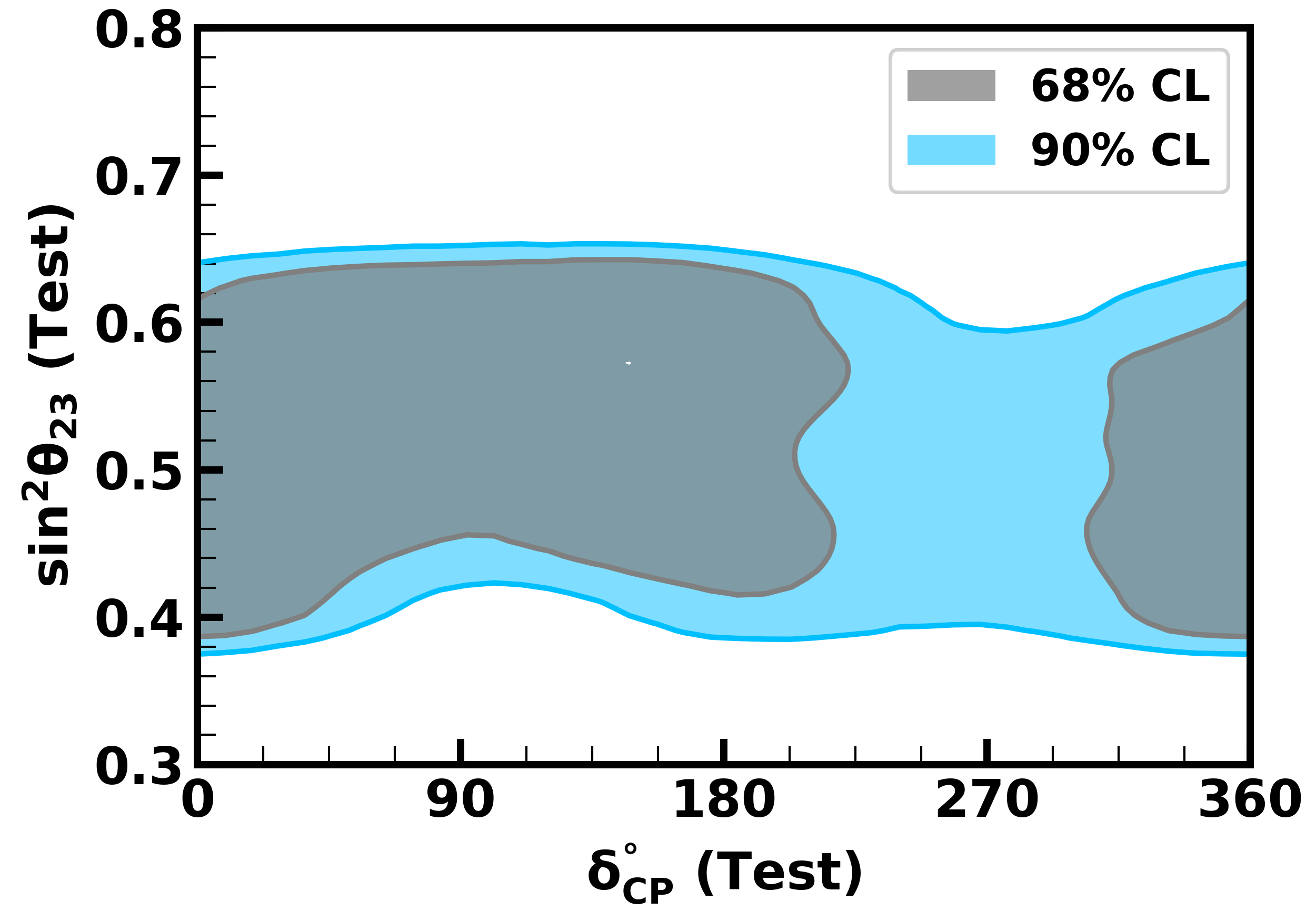}
\includegraphics[width=0.45\linewidth]{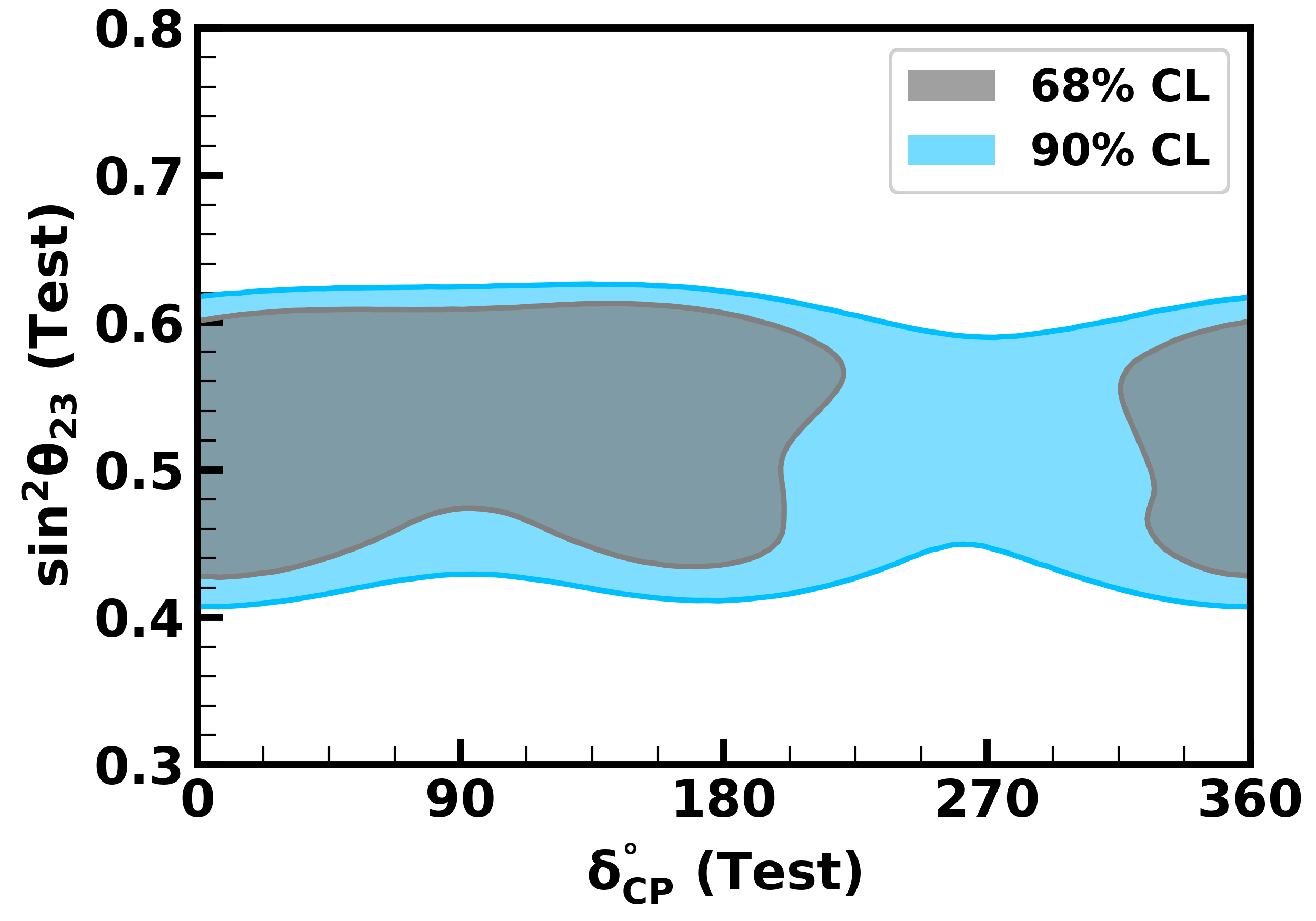}
\includegraphics[width=0.45\linewidth]{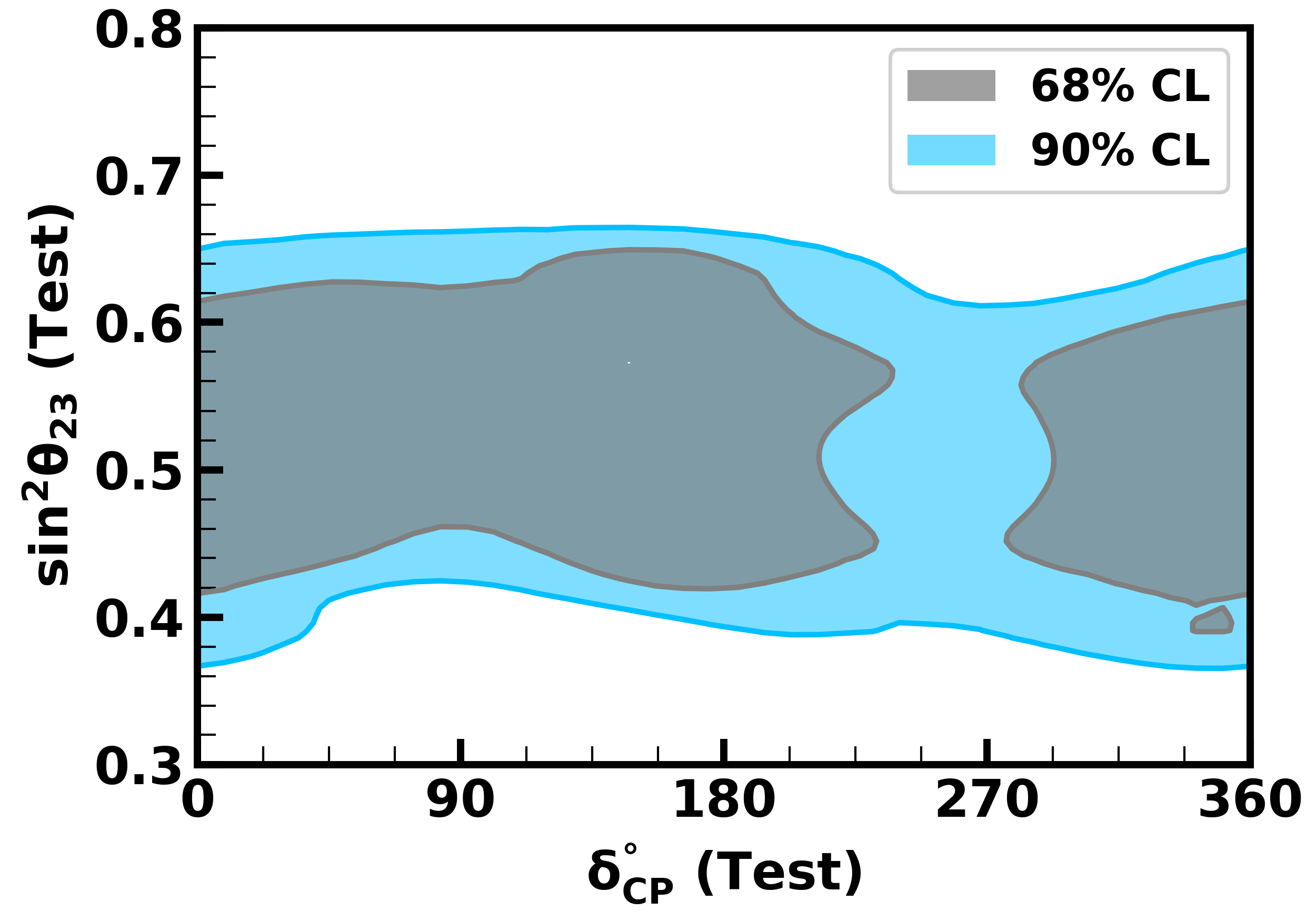}
\includegraphics[width=0.45\linewidth]{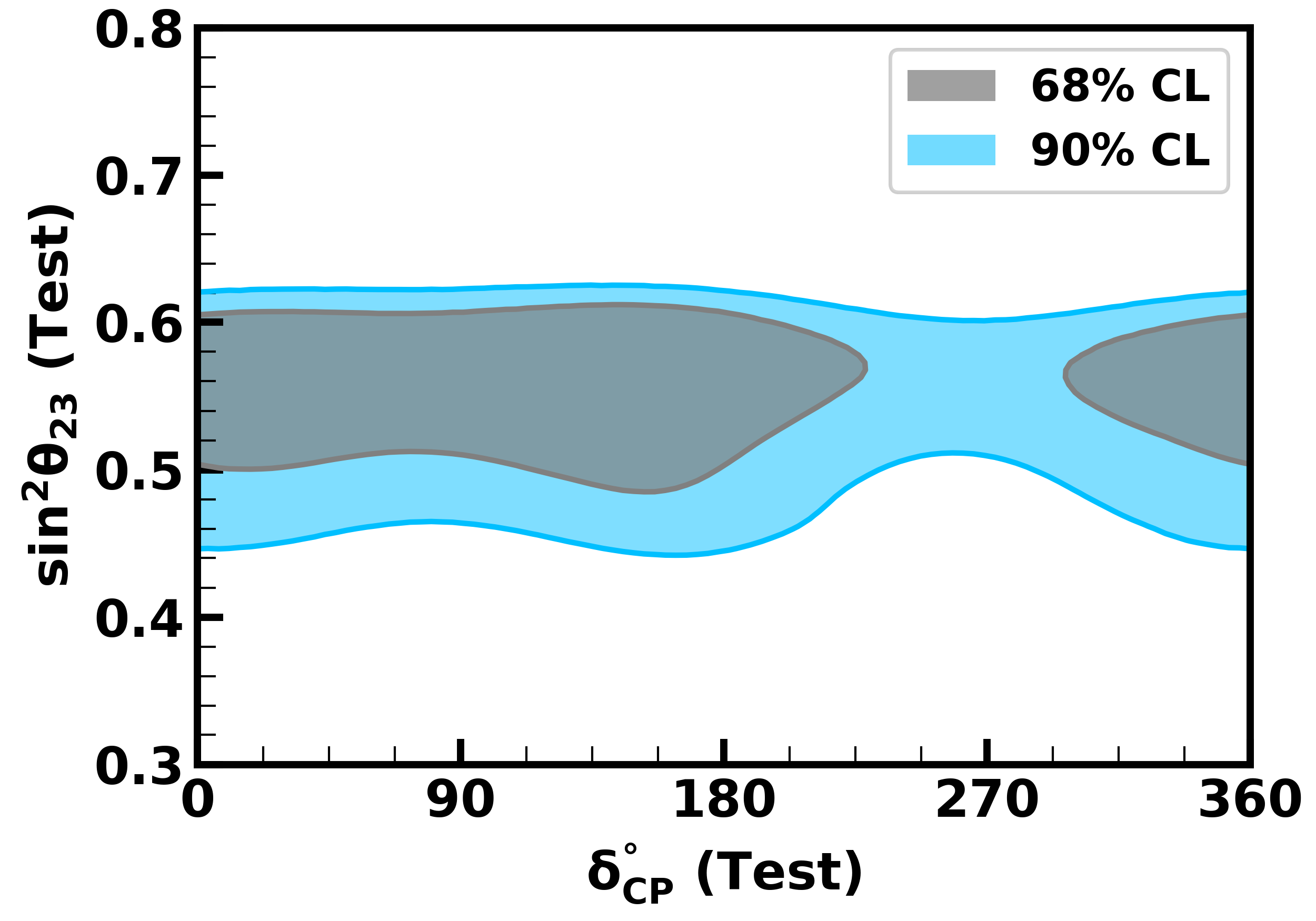}
\caption{$\Gamma_{21} = \Gamma_{31} = \Gamma_{32} = 10^{-23}$ GeV in true for $n = 0$ in top row, $n=1$ in middle row and $n=2$ in bottom row}. In left 1 - 5 GeV and in right 1 - 20 GeV. Marginalized over $\Delta m_{31}^2$, $\theta_{13}$ and $\Gamma$
\label{fig:theta23-delcp}
\end{figure*}

In this sub-section we choose to highlight the most stringent bounds on $\theta_{23}$ and $\delta_{CP}$ imposed by assuming non-zero $\Gamma$ with a power-law index $n = 2$. We achieve this by assuming decoherence in both the simulated data and the test hypothesis. To obtain the the projection of $\chi^2$ in then $\theta_{23} - \delta_{CP}$ test plane we define $\chi^2$ as follows
\begin{equation}
    \chi^2 = \chi^2(\Gamma^{true},\delta^{true}_{CP},\theta^{true}_{23},\delta^{test}_{CP},\theta^{test}_{23})~.
\end{equation}
We marginalize over test $\Delta m_{31}^2$, $\theta_{13}$ and $\Gamma~[10^{-24}:10^{-22}]$ GeV. We consider true values of standard parameters and their test ranges from table~\ref{table:1} assuming normal mass ordering. We fix the unmentioned parameters to their true values.

In fig.~\ref{fig:theta23-delcp} we display the projection of $\chi^2$ in the test plane of $\theta_{23} - \delta_{CP}$ plane considering true $\Gamma_0 = 10^{-23}~GeV$ (a representative value) and energy power-law indices $n = 0,~1$ and $2$ in the top, middle and bottom panel. The left and right plots represent allowed regions considering analysis window of neutrino energy 1 to 4 GeV and 1 to 20 GeV respectively. In both the plots we show the contours corresponding to $68\%$ (gray) and $90\%$ (blue) CL.

From the left plot of fig.~\ref{fig:theta23-delcp} ($E_\nu : 1-4$ GeV) we notice that even in the presence of decoherence the allowed regions of $\theta_{23}$ and $\delta_{CP}$ are almost the same as in the standard case reported in~\cite{NOvA:2021nfi}. This implies that when we consider the $\nu_e$ events in the range $E_\nu : 1-4$ GeV the impact of the presence of decoherence, in true spectrum is very minimal (even for power law index $n=2$). On the contrary, when we considered $E_\nu : 4-20$~GeV (right panel), we note that the allowed regions of $\theta_{23}$ for all values of $\delta_{CP}$ got more stringent for $n=1,2$ and the higher octant of $\theta_{23}$ ($\theta_{23} > 45\degree$) is preferred over the lower octant ($\theta_{23} < 45\degree$) for $n=2$.

Furthermore, we observe a disfavor region for $\delta_{CP} \sim 270\degree$ with $68\%$ CL in both the left and right plots even in the presence of decoherence and the region is allowed with $90\%$ CL.
Therefore, in the presence of decoherence the NOvA and T2K tension shown in ref.~\cite{NOvA:2021nfi} (normal ordering plot in fig. 6), could be uplifted with $90\%$ CL.

\section{Conclusions}\label{sec:conclusion}
In this paper we have studied the phenomenology of NSI and environmental decoherence by including high energy neutrino events at NOvA experiment. We have shown the modified oscillation probabilities and event rates at high energy considering effect of one NSI parameters ($\epsilon_{e\mu}$, $\epsilon_{e\tau}$ at a time) as well as decoherence parameter $\Gamma$. We have illustrated the allowed regions with $95\%$ CL in the $|\epsilon_{\alpha\beta}| - \delta_{CP(\alpha\beta)}$ plane taking into account high energy $\nu_e$ and $\bar{\nu}_e$ events. We have obtained the bounds on decoherence parameter $\Gamma$ considering the power-law dependencies $n=0,1,2$. In the decoherence scenario we have highlighted the bounds on $\theta_{23}$ and $\delta_{CP}$ imposed by assuming power-law indices $n = 0,1,2$ for the analysis window $E_\nu : 1-20$ GeV.

The standard appearance probability and the probability with non-zero $\epsilon_{e\tau}$ exhibit the degeneracy throughout the energy window considered in this analysis. Interestingly, the degeneracy observed for higher $\epsilon_{e\tau}$ in the $|\epsilon_{e\tau}|-\delta_{CP(e\tau)}$ plane disappears if we include events from high energy range ($E_\nu: 4-20$ GeV). On the other hand the appearance probability increases at high energy ($E_\nu > 7$ GeV) as compared to standard oscillation probability in the presence of $\epsilon_{e\mu}$ and we do not observe degeneracy with the standard oscillation probability. In addition, the $95\%$ CL contours in $|\epsilon_{e\mu}| - \delta_{CP(e\mu)}$ plane show that $\epsilon_{e\mu}$ can be tightly constrained for all values of $\delta_{CP}$ after including high energy events.

In the case of decoherence we have seen an increase in $\nu_e$-appearance probability and events $\sim 11$ GeV for $n \geq 0$. Including the high energy events we have obtained stringent upper bounds on $\Gamma$ specifically for power-law indices $n=1,~2$. Considering non-zero $\Gamma$ with $n=2$, we have shown that $\theta_{23}$ is better constrained for all values of $\delta_{CP}$ and the higher octant of $\theta_{23}$ ($\theta_{23} > 45\degree$) is preferred over the lower octant ($\theta_{23} < 45\degree$) including events from $E_\nu : 4-20$~GeV. 

\noindent{\bf Acknowledgments}

\noindent We would like to thank Prof. Mark Messier and Prof. Jon Urheim, Indiana University, USA, for their continuous assistance and valuable support with this work. Special thanks to Ishwar Singh, University of Delhi, for fruitful discussions. 
One of the authors (K. N. D.) acknowledges the Department of Science and Technology (DST)-SERB International Research Experience (SIRE) program for the financial support Grant No. SIR/2022/000518 to visit Indiana University, USA. We acknowledge high performance computing (HPC) facilities in \'Ecole Centrale School of Engineering - Mahindra University.

\appendix
\section{$\epsilon_{e \tau}$ vs $\delta_{CP}$ ($\delta_{e\tau}$) assuming no signal in $E_\nu: 5-20$ GeV}\label{app:eps_etau_delcp-no_sig}
\begin{figure*}[!htbp]
\includegraphics[width=0.45\linewidth]{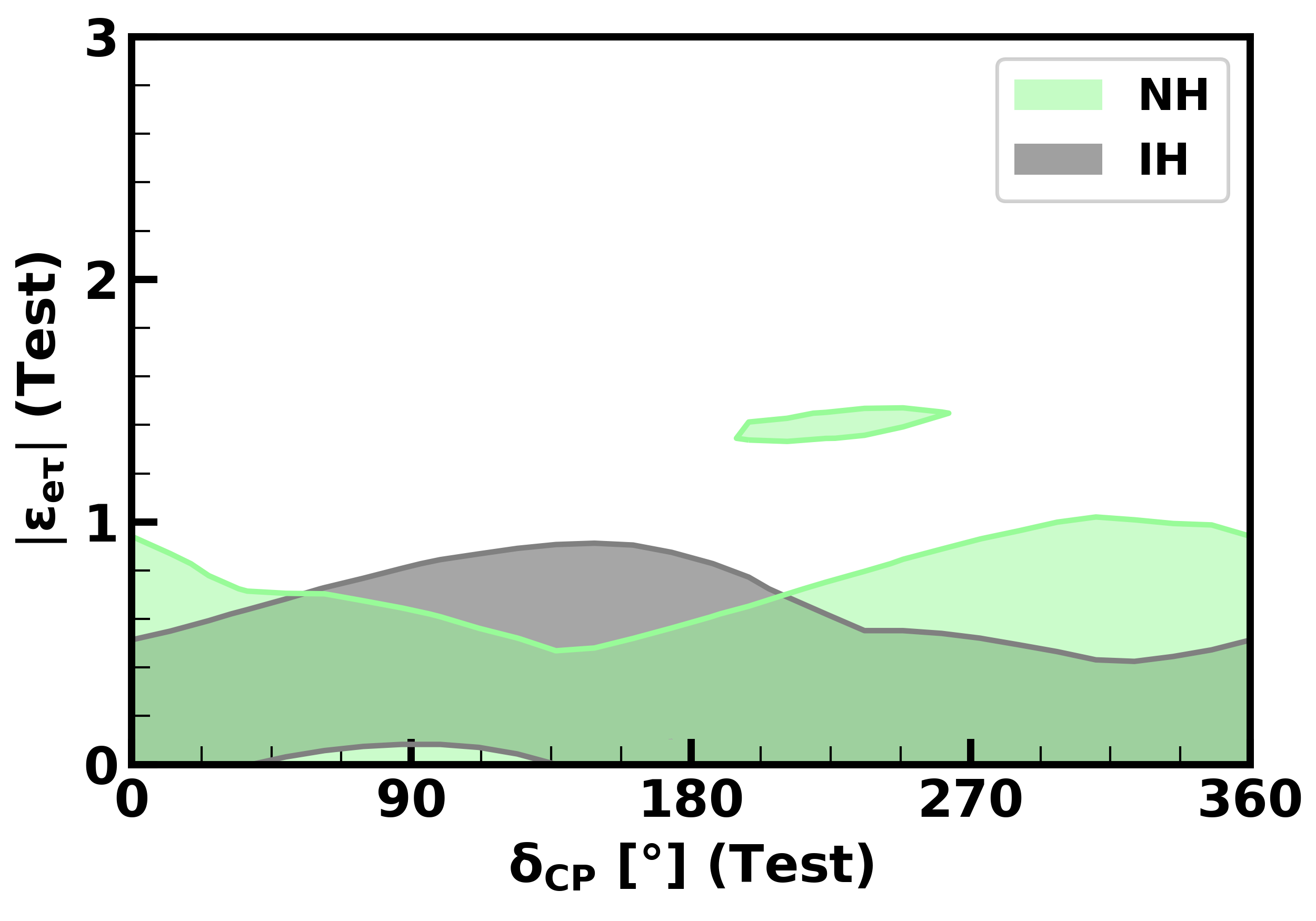}
\includegraphics[width=0.45\linewidth]{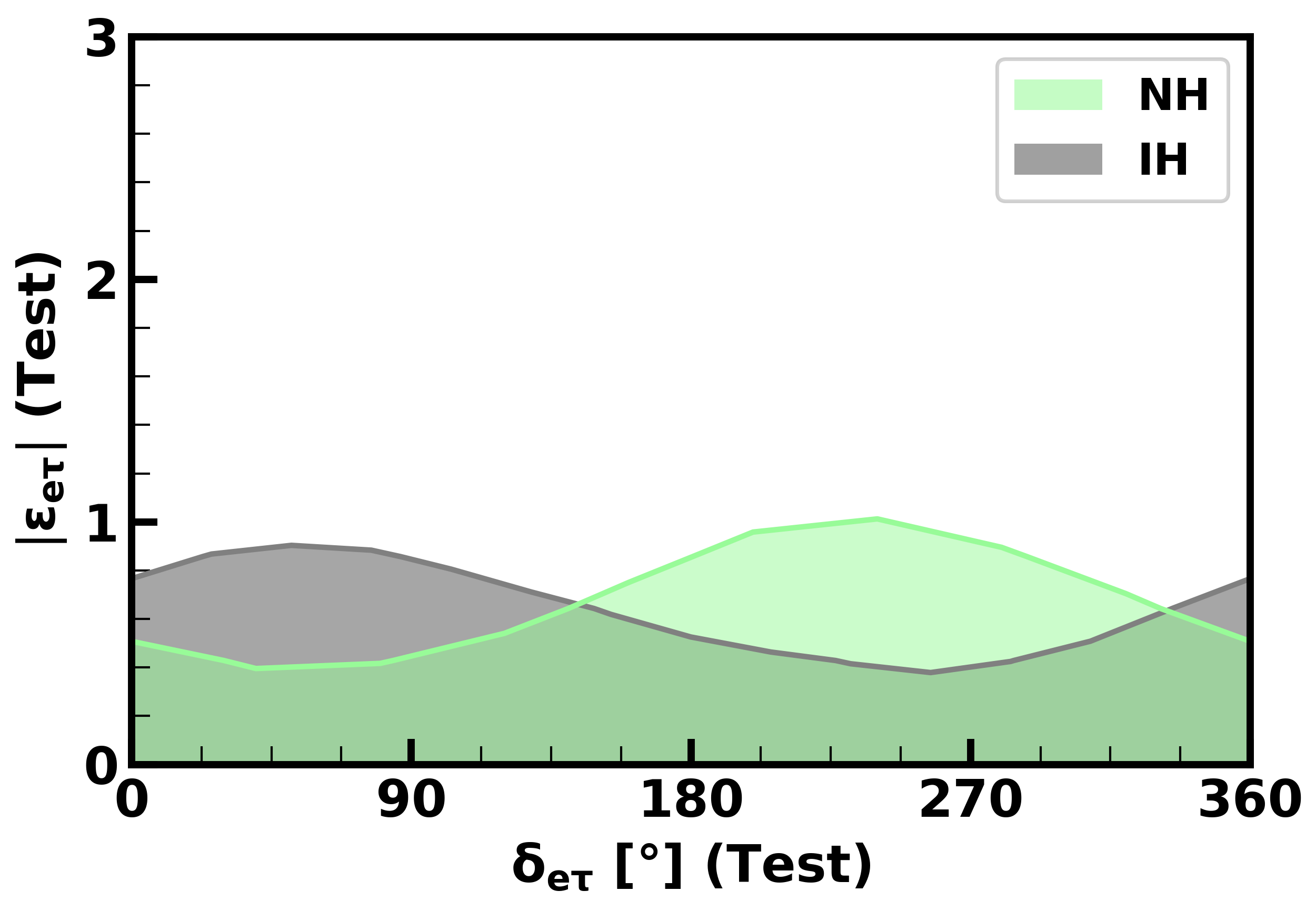}
\caption{$\epsilon_{e \tau}$ vs $\delta_{CP}$ (left) and $\epsilon_{e \tau}$ vs $\delta_{e\tau}$ (right) for only background (no signal) in energy range $E_\nu: 5-20$ GeV. }
\label{fig:eps_etau-delcp_deletau-no_sig-5-20GeV}
\end{figure*}

In fig.~\ref{fig:eps_etau-delcp_deletau-no_sig-5-20GeV}, we consider simulated data from $E_\nu: 1-4$ GeV accompany signal and background, whereas $E_\nu: 4-20$ GeV spectra accompany only background.


\FloatBarrier
\bibliographystyle{elsarticle-num}
\bibliography{bibliography.bib}

\begin{thebibliography}{10}
\expandafter\ifx\csname url\endcsname\relax
  \def\url#1{\texttt{#1}}\fi
\expandafter\ifx\csname urlprefix\endcsname\relax\def\urlprefix{URL }\fi
\expandafter\ifx\csname href\endcsname\relax
  \def\href#1#2{#2} \def\path#1{#1}\fi

\bibitem{Wolfenstein:1977ue}
L.~Wolfenstein, {Neutrino Oscillations in Matter}, Phys. Rev. D 17 (1978) 2369--2374.
\newblock \href {https://doi.org/10.1103/PhysRevD.17.2369} {\path{doi:10.1103/PhysRevD.17.2369}}.

\bibitem{Roulet:1991sm}
E.~Roulet, {MSW effect with flavor changing neutrino interactions}, Phys. Rev. D 44 (1991) R935--R938.
\newblock \href {https://doi.org/10.1103/PhysRevD.44.R935} {\path{doi:10.1103/PhysRevD.44.R935}}.

\bibitem{Hattori:2002uw}
T.~Hattori, T.~Hasuike, S.~Wakaizumi, {Flavor changing neutrino interactions and CP violation in neutrino oscillations}, Prog. Theor. Phys. 114 (2005) 439--449.
\newblock \href {http://arxiv.org/abs/hep-ph/0210138} {\path{arXiv:hep-ph/0210138}}, \href {https://doi.org/10.1143/PTP.114.439} {\path{doi:10.1143/PTP.114.439}}.

\bibitem{Chang:1998ea}
C.-H. Chang, W.-S. Dai, X.-Q. Li, Y.~Liu, F.-C. Ma, Z.-j. Tao, {Possible effects of quantum mechanics violation induced by certain quantum gravity on neutrino oscillations}, Phys. Rev. D 60 (1999) 033006.
\newblock \href {http://arxiv.org/abs/hep-ph/9809371} {\path{arXiv:hep-ph/9809371}}, \href {https://doi.org/10.1103/PhysRevD.60.033006} {\path{doi:10.1103/PhysRevD.60.033006}}.

\bibitem{Benatti:2000ph}
F.~Benatti, R.~Floreanini, {Open system approach to neutrino oscillations}, JHEP 02 (2000) 032.
\newblock \href {http://arxiv.org/abs/hep-ph/0002221} {\path{arXiv:hep-ph/0002221}}, \href {https://doi.org/10.1088/1126-6708/2000/02/032} {\path{doi:10.1088/1126-6708/2000/02/032}}.

\bibitem{Farzan:2008zv}
Y.~Farzan, T.~Schwetz, A.~Y. Smirnov, {Reconciling results of LSND, MiniBooNE and other experiments with soft decoherence}, JHEP 07 (2008) 067.
\newblock \href {http://arxiv.org/abs/0805.2098} {\path{arXiv:0805.2098}}, \href {https://doi.org/10.1088/1126-6708/2008/07/067} {\path{doi:10.1088/1126-6708/2008/07/067}}.

\bibitem{Denton:2020uda}
P.~B. Denton, J.~Gehrlein, R.~Pestes, {$CP$ -Violating Neutrino Nonstandard Interactions in Long-Baseline-Accelerator Data}, Phys. Rev. Lett. 126~(5) (2021) 051801.
\newblock \href {http://arxiv.org/abs/2008.01110} {\path{arXiv:2008.01110}}, \href {https://doi.org/10.1103/PhysRevLett.126.051801} {\path{doi:10.1103/PhysRevLett.126.051801}}.

\bibitem{NOvA:2021nfi}
M.~A. Acero, et~al., {Improved measurement of neutrino oscillation parameters by the NOvA experiment}, Phys. Rev. D 106~(3) (2022) 032004.
\newblock \href {http://arxiv.org/abs/2108.08219} {\path{arXiv:2108.08219}}, \href {https://doi.org/10.1103/PhysRevD.106.032004} {\path{doi:10.1103/PhysRevD.106.032004}}.

\bibitem{T2K:2021xwb}
K.~Abe, et~al., {Improved constraints on neutrino mixing from the T2K experiment with $\mathbf{3.13\times10^{21}}$ protons on target}, Phys. Rev. D 103~(11) (2021) 112008.
\newblock \href {http://arxiv.org/abs/2101.03779} {\path{arXiv:2101.03779}}, \href {https://doi.org/10.1103/PhysRevD.103.112008} {\path{doi:10.1103/PhysRevD.103.112008}}.

\bibitem{Denton:2022pxt}
P.~B. Denton, A.~Giarnetti, D.~Meloni, {How to identify different new neutrino oscillation physics scenarios at DUNE}, JHEP 02 (2023) 210.
\newblock \href {http://arxiv.org/abs/2210.00109} {\path{arXiv:2210.00109}}, \href {https://doi.org/10.1007/JHEP02(2023)210} {\path{doi:10.1007/JHEP02(2023)210}}.

\bibitem{NOvA:2024lti}
M.~A. Acero, et~al., {Search for CP-Violating Neutrino Nonstandard Interactions with the NOvA Experiment}, Phys. Rev. Lett. 133~(20) (2024) 201802.
\newblock \href {http://arxiv.org/abs/2403.07266} {\path{arXiv:2403.07266}}, \href {https://doi.org/10.1103/PhysRevLett.133.201802} {\path{doi:10.1103/PhysRevLett.133.201802}}.

\bibitem{Kleykamp:2022dli}
J.~Kleykamp, L.~R. Prais, M.~A. Acero~Ortega, G.~S. Davies, {Non-standard interactions at NOvA} (2022).
\newblock \href {https://doi.org/10.2172/1874281} {\path{doi:10.2172/1874281}}.

\bibitem{Coelho:2017zes}
J.~A.~B. Coelho, W.~A. Mann, S.~S. Bashar, {Nonmaximal $\theta_{23}$ mixing at NOvA from neutrino decoherence}, Phys. Rev. Lett. 118~(22) (2017) 221801.
\newblock \href {http://arxiv.org/abs/1702.04738} {\path{arXiv:1702.04738}}, \href {https://doi.org/10.1103/PhysRevLett.118.221801} {\path{doi:10.1103/PhysRevLett.118.221801}}.

\bibitem{DeRomeri:2023dht}
V.~De~Romeri, C.~Giunti, T.~Stuttard, C.~A. Ternes, {Neutrino oscillation bounds on quantum decoherence}, JHEP 09 (2023) 097.
\newblock \href {http://arxiv.org/abs/2306.14699} {\path{arXiv:2306.14699}}, \href {https://doi.org/10.1007/JHEP09(2023)097} {\path{doi:10.1007/JHEP09(2023)097}}.

\bibitem{2735647}
I.~Singh, P.~Singh, B.~Choudhary, {Impact of High Energy \ensuremath{\nu}e (\ensuremath{\nu}\textasciimacron{}e) Events on NOvA Oscillation Sensitivities}, Springer Proceedings in Physics 304 (2024) 174--178.
\newblock \href {https://doi.org/10.1007/978-981-97-0289-3_37} {\path{doi:10.1007/978-981-97-0289-3_37}}.

\bibitem{Shanahan:2021jlp}
P.~N. Shanahan, P.~L. Vahle, {Physics with NOvA: a half-time review}, Eur. Phys. J. ST 230~(24) (2021) 4259--4273.
\newblock \href {https://doi.org/10.1140/epjs/s11734-021-00285-9} {\path{doi:10.1140/epjs/s11734-021-00285-9}}.

\bibitem{Huber:2004ka}
P.~Huber, M.~Lindner, W.~Winter, {Simulation of long-baseline neutrino oscillation experiments with GLoBES (General Long Baseline Experiment Simulator)}, Comput. Phys. Commun. 167 (2005) 195.
\newblock \href {http://arxiv.org/abs/hep-ph/0407333} {\path{arXiv:hep-ph/0407333}}, \href {https://doi.org/10.1016/j.cpc.2005.01.003} {\path{doi:10.1016/j.cpc.2005.01.003}}.

\bibitem{Huber:2007ji}
P.~Huber, J.~Kopp, M.~Lindner, M.~Rolinec, W.~Winter, {New features in the simulation of neutrino oscillation experiments with GLoBES 3.0: General Long Baseline Experiment Simulator}, Comput. Phys. Commun. 177 (2007) 432--438.
\newblock \href {http://arxiv.org/abs/hep-ph/0701187} {\path{arXiv:hep-ph/0701187}}, \href {https://doi.org/10.1016/j.cpc.2007.05.004} {\path{doi:10.1016/j.cpc.2007.05.004}}.

\bibitem{NOvA:2018gge}
M.~A. Acero, et~al., {New constraints on oscillation parameters from $\nu_e$ appearance and $\nu_\mu$ disappearance in the NOvA experiment}, Phys. Rev. D 98 (2018) 032012.
\newblock \href {http://arxiv.org/abs/1806.00096} {\path{arXiv:1806.00096}}, \href {https://doi.org/10.1103/PhysRevD.98.032012} {\path{doi:10.1103/PhysRevD.98.032012}}.

\bibitem{NOvA:2019cyt}
M.~A. Acero, et~al., {First Measurement of Neutrino Oscillation Parameters using Neutrinos and Antineutrinos by NOvA}, Phys. Rev. Lett. 123~(15) (2019) 151803.
\newblock \href {http://arxiv.org/abs/1906.04907} {\path{arXiv:1906.04907}}, \href {https://doi.org/10.1103/PhysRevLett.123.151803} {\path{doi:10.1103/PhysRevLett.123.151803}}.

\bibitem{Kopp:2006wp}
J.~Kopp, {Efficient numerical diagonalization of hermitian 3 x 3 matrices}, Int. J. Mod. Phys. C 19 (2008) 523--548.
\newblock \href {http://arxiv.org/abs/physics/0610206} {\path{arXiv:physics/0610206}}, \href {https://doi.org/10.1142/S0129183108012303} {\path{doi:10.1142/S0129183108012303}}.

\bibitem{Kopp:2007ne}
J.~Kopp, M.~Lindner, T.~Ota, J.~Sato, {Non-standard neutrino interactions in reactor and superbeam experiments}, Phys. Rev. D 77 (2008) 013007.
\newblock \href {http://arxiv.org/abs/0708.0152} {\path{arXiv:0708.0152}}, \href {https://doi.org/10.1103/PhysRevD.77.013007} {\path{doi:10.1103/PhysRevD.77.013007}}.

\bibitem{Fogli:2002pt}
G.~L. Fogli, E.~Lisi, A.~Marrone, D.~Montanino, A.~Palazzo, {Getting the most from the statistical analysis of solar neutrino oscillations}, Phys. Rev. D 66 (2002) 053010.
\newblock \href {http://arxiv.org/abs/hep-ph/0206162} {\path{arXiv:hep-ph/0206162}}, \href {https://doi.org/10.1103/PhysRevD.66.053010} {\path{doi:10.1103/PhysRevD.66.053010}}.

\bibitem{Huber:2002mx}
P.~Huber, M.~Lindner, W.~Winter, {Superbeams versus neutrino factories}, Nucl. Phys. B 645 (2002) 3--48.
\newblock \href {http://arxiv.org/abs/hep-ph/0204352} {\path{arXiv:hep-ph/0204352}}, \href {https://doi.org/10.1016/S0550-3213(02)00825-8} {\path{doi:10.1016/S0550-3213(02)00825-8}}.

\bibitem{ParticleDataGroup:2018ovx}
M.~Tanabashi, et~al., {Review of Particle Physics}, Phys. Rev. D 98~(3) (2018) 030001.
\newblock \href {https://doi.org/10.1103/PhysRevD.98.030001} {\path{doi:10.1103/PhysRevD.98.030001}}.

\bibitem{Farzan:2017xzy}
Y.~Farzan, M.~Tortola, {Neutrino oscillations and Non-Standard Interactions}, Front. in Phys. 6 (2018) 10.
\newblock \href {http://arxiv.org/abs/1710.09360} {\path{arXiv:1710.09360}}, \href {https://doi.org/10.3389/fphy.2018.00010} {\path{doi:10.3389/fphy.2018.00010}}.

\bibitem{Biggio:2009nt}
C.~Biggio, M.~Blennow, E.~Fernandez-Martinez, {General bounds on non-standard neutrino interactions}, JHEP 08 (2009) 090.
\newblock \href {http://arxiv.org/abs/0907.0097} {\path{arXiv:0907.0097}}, \href {https://doi.org/10.1088/1126-6708/2009/08/090} {\path{doi:10.1088/1126-6708/2009/08/090}}.

\bibitem{Ohlsson:2012kf}
T.~Ohlsson, {Status of non-standard neutrino interactions}, Rept. Prog. Phys. 76 (2013) 044201.
\newblock \href {http://arxiv.org/abs/1209.2710} {\path{arXiv:1209.2710}}, \href {https://doi.org/10.1088/0034-4885/76/4/044201} {\path{doi:10.1088/0034-4885/76/4/044201}}.

\bibitem{Miranda:2015dra}
O.~G. Miranda, H.~Nunokawa, {Non standard neutrino interactions: current status and future prospects}, New J. Phys. 17~(9) (2015) 095002.
\newblock \href {http://arxiv.org/abs/1505.06254} {\path{arXiv:1505.06254}}, \href {https://doi.org/10.1088/1367-2630/17/9/095002} {\path{doi:10.1088/1367-2630/17/9/095002}}.

\bibitem{Proceedings:2019qno}
{Neutrino Non-Standard Interactions: A Status Report}, Vol.~2.
\newblock \href {http://arxiv.org/abs/1907.00991} {\path{arXiv:1907.00991}}, \href {https://doi.org/10.21468/SciPostPhysProc.2.001} {\path{doi:10.21468/SciPostPhysProc.2.001}}.

\bibitem{Bera:2025ayt}
C.~Bera, K.~N. Deepthi, R.~Mohanta, {The effect of non-standard interactions and environmental decoherence at DUNE} (1 2025).
\newblock \href {http://arxiv.org/abs/2501.14383} {\path{arXiv:2501.14383}}.

\bibitem{Stuttard:2020qfv}
T.~Stuttard, M.~Jensen, {Neutrino decoherence from quantum gravitational stochastic perturbations}, Phys. Rev. D 102~(11) (2020) 115003.
\newblock \href {http://arxiv.org/abs/2007.00068} {\path{arXiv:2007.00068}}, \href {https://doi.org/10.1103/PhysRevD.102.115003} {\path{doi:10.1103/PhysRevD.102.115003}}.

\bibitem{Giunti:1998kim}
C.~Giunti, C.~W. Kim, {Coherence of neutrino oscillations in the wave packet approach}, Phys. Rev. D 58 (1998) 017301.
\newblock \href {http://arxiv.org/abs/hep-ph/9711363} {\path{arXiv:hep-ph/9711363}}, \href {https://doi.org/10.1103/PhysRevD.58.017301} {\path{doi:10.1103/PhysRevD.58.017301}}.

\bibitem{Blennow:2005ohl}
M.~Blennow, T.~Ohlsson, W.~Winter, {Damping signatures in future neutrino oscillation experiments}, JHEP 06 (2005) 049.
\newblock \href {http://arxiv.org/abs/hep-ph/0502147} {\path{arXiv:hep-ph/0502147}}, \href {https://doi.org/10.1088/1126-6708/2005/06/049} {\path{doi:10.1088/1126-6708/2005/06/049}}.

\bibitem{Akhmedov:2012her}
E.~Akhmedov, D.~Hernandez, A.~Smirnov, {Neutrino production coherence and oscillation experiments}, JHEP 04 (2012) 052.
\newblock \href {http://arxiv.org/abs/1201.4128} {\path{arXiv:1201.4128}}, \href {https://doi.org/10.1007/JHEP04(2012)052} {\path{doi:10.1007/JHEP04(2012)052}}.

\bibitem{Chang:2016chu}
Y.-L. Chan, M.~C. Chu, K.~M. Tsui, C.~F. Wong, J.~Xu, {Wave-packet treatment of reactor neutrino oscillation experiments and its implications on determining the neutrino mass hierarchy}, Eur. Phys. J. C 76~(6) (2016) 310.
\newblock \href {http://arxiv.org/abs/1507.06421} {\path{arXiv:1507.06421}}, \href {https://doi.org/10.1140/epjc/s10052-016-4143-4} {\path{doi:10.1140/epjc/s10052-016-4143-4}}.

\bibitem{Gouvea:2021rom}
A.~de~Gouv\^ea, V.~De~Romeri, C.~A. Ternes, {Combined analysis of neutrino decoherence at reactor experiments}, JHEP 06 (2021) 042.
\newblock \href {http://arxiv.org/abs/2104.05806} {\path{arXiv:2104.05806}}, \href {https://doi.org/10.1007/JHEP06(2021)042} {\path{doi:10.1007/JHEP06(2021)042}}.

\bibitem{Lindblad:1976g}
G.~Lindblad, {On the Generators of Quantum Dynamical Semigroups}, Commun. Math. Phys. 48 (1976) 119.
\newblock \href {https://doi.org/10.1007/BF01608499} {\path{doi:10.1007/BF01608499}}.

\bibitem{GKS:1976vit}
V.~Gorini, A.~Kossakowski, E.~C.~G. Sudarshan, {Completely Positive Dynamical Semigroups of N Level Systems}, J. Math. Phys. 17 (1976) 821.
\newblock \href {https://doi.org/10.1063/1.522979} {\path{doi:10.1063/1.522979}}.

\bibitem{GKS:1978g}
V.~Gorini, A.~Frigerio, M.~Verri, A.~Kossakowski, E.~C.~G. Sudarshan, {Properties of Quantum Markovian Master Equations}, Rept. Math. Phys. 13 (1978) 149.
\newblock \href {https://doi.org/10.1016/0034-4877(78)90050-2} {\path{doi:10.1016/0034-4877(78)90050-2}}.

\bibitem{Banks:1984bsp}
T.~Banks, L.~Susskind, M.~E. Peskin, {Difficulties for the Evolution of Pure States Into Mixed States}, Nucl. Phys. B 244 (1984) 125--134.
\newblock \href {https://doi.org/10.1016/0550-3213(84)90184-6} {\path{doi:10.1016/0550-3213(84)90184-6}}.

\bibitem{Benatti:1988nar}
F.~Benatti, H.~Narnhofer, {ENTROPY BEHAVIOR UNDER COMPLETELY POSITIVE MAPS}, Lett. Math. Phys. 15 (1988) 325.
\newblock \href {https://doi.org/10.1007/BF00419590} {\path{doi:10.1007/BF00419590}}.

\bibitem{Denton:2018dmp}
P.~B. Denton, S.~J. Parke, {Addendum to ``Compact perturbative expressions for neutrino oscillations in matter''}[Addendum: JHEP 06, 109 (2018)] (1 2018).
\newblock \href {http://arxiv.org/abs/1801.06514} {\path{arXiv:1801.06514}}, \href {https://doi.org/10.1007/JHEP06(2018)109} {\path{doi:10.1007/JHEP06(2018)109}}.

\bibitem{Gomes:2019for}
G.~Balieiro~Gomes, D.~V. Forero, M.~M. Guzzo, P.~C. De~Holanda, R.~L.~N. Oliveira, {Quantum Decoherence Effects in Neutrino Oscillations at DUNE}, Phys. Rev. D 100~(5) (2019) 055023.
\newblock \href {http://arxiv.org/abs/1805.09818} {\path{arXiv:1805.09818}}, \href {https://doi.org/10.1103/PhysRevD.100.055023} {\path{doi:10.1103/PhysRevD.100.055023}}.

\bibitem{Coloma:2018ice}
P.~Coloma, J.~Lopez-Pavon, I.~Martinez-Soler, H.~Nunokawa, {Decoherence in Neutrino Propagation Through Matter, and Bounds from IceCube/DeepCore}, Eur. Phys. J. C 78~(8) (2018) 614.
\newblock \href {http://arxiv.org/abs/1803.04438} {\path{arXiv:1803.04438}}, \href {https://doi.org/10.1140/epjc/s10052-018-6092-6} {\path{doi:10.1140/epjc/s10052-018-6092-6}}.

\bibitem{Romeri:2023cgt}
V.~De~Romeri, C.~Giunti, T.~Stuttard, C.~A. Ternes, {Neutrino oscillation bounds on quantum decoherence}, JHEP 09 (2023) 097.
\newblock \href {http://arxiv.org/abs/2306.14699} {\path{arXiv:2306.14699}}, \href {https://doi.org/10.1007/JHEP09(2023)097} {\path{doi:10.1007/JHEP09(2023)097}}.

\bibitem{Bera:2024hhr}
C.~Bera, K.~N. Deepthi, {Study of quantum decoherence at the Protvino to ORCA experiment}, Phys. Rev. D 110~(3) (2024) 035035.
\newblock \href {http://arxiv.org/abs/2405.03286} {\path{arXiv:2405.03286}}, \href {https://doi.org/10.1103/PhysRevD.110.035035} {\path{doi:10.1103/PhysRevD.110.035035}}.

\bibitem{Ellis:1999uh}
J.~R. Ellis, N.~E. Mavromatos, D.~V. Nanopoulos, {Quantum gravitational diffusion and stochastic fluctuations in the velocity of light}, Gen. Rel. Grav. 32 (2000) 127--144.
\newblock \href {http://arxiv.org/abs/gr-qc/9904068} {\path{arXiv:gr-qc/9904068}}, \href {https://doi.org/10.1023/A:1001852601248} {\path{doi:10.1023/A:1001852601248}}.

\bibitem{Ellis:1997jw}
J.~R. Ellis, N.~E. Mavromatos, D.~V. Nanopoulos, {Quantum decoherence in a D foam background}, Mod. Phys. Lett. A 12 (1997) 1759--1773.
\newblock \href {http://arxiv.org/abs/hep-th/9704169} {\path{arXiv:hep-th/9704169}}, \href {https://doi.org/10.1142/S0217732397001795} {\path{doi:10.1142/S0217732397001795}}.

\end{thebibliography}

\end{document}